\documentstyle[amssymb,aps]{revtex}

\begin{document}
\draft
\title{Chaos in a double driven dissipative nonlinear oscillator}
\author{H. H. Adamyan, S. B. Manvelyan, and G. Yu. Kryuchkyan}
\address{Yerevan State University, Manookyan 1, Yerevan, \ 375049, Armenia\\
Institute for Physical Research, National Academy of Sciences, Ashtarak-2,\\
378410, Armenia}
\maketitle

\begin{abstract}
We propose an anharmonic oscillator driven by two periodic forces of
different frequencies as a new time-dependent model for investigating
quantum dissipative chaos. Our analysis is done in the frame of statistical
ensemble of quantum trajectories in quantum state diffusion approach.
Quantum dynamical manifestation of chaotic behavior, including the emergence
of chaos, properties of strange attractors, and quantum entanglement are
studied by numerical simulation of ensemble averaged Wigner function and von
Neumann entropy.
\end{abstract}


\section{ Introduction}

Quantum nonlinear systems with chaotic classical counterparts have received
much attention in the last two decades. This field of investigation is
sometimes called ''Quantum Chaos'' \cite{1}. The usual procedure of studying
a quantum chaos is to take a system which exhibits chaotic motion,when
treated classically, and see what effects occur in a quantum treatment. All
real quantum systems are open and their classical limit is related to the
loss of coherence produced by interaction with the environment. Thus,
investigations of quantum chaotic system are connected with the
correspondence problem in general, and with decoherence and dissipation in
particular.

It is now well established that the quantum dynamics of classically chaotic
systems will show major departures from the classical motion on a suitable
time scale. Among these phenomena we note the dynamical localization of
classical diffusive excitation, due to quantum mechanical interference,
which is in close analogy with the Anderson localization in a random
potential. Dynamical localization has been well studied theoretically \cite
{2} and verified in experiments \cite{3} with laser-cooled atom moving in a
standing wave with periodically modulated nodal position. Much research on
the subject of classical and quantum chaos is devoted to the kicked rotator,
which exhibits regions of regular and chaotic motion in the Poincar\'{e}
section (see for example \cite{4}). This model is very popular in
investigations of transition to quantum chaos. Its experimental realization,
and observation of the model's dissipation and decoherence effects,\ is
carried out on gas of ultracold atoms in a magneto-optical trap subjected to
a pulsed standing wave \cite{5,6}. In Ref.\cite{7} it was proposed to
realize the parametrically kicked nonlinear oscillator model in a cavity
involving Kerr nonlinearity. It was also shown that more promising
realization of this system also including quantum regime, is achieved in the
dynamics of cooled and trapped ion, interacting with periodic sequence of
both standing wave pulses and Gaussian laser pulses \cite{8}. Another
suggestion to investigate quantum chaos in a single trapped ion was recently
provided in \cite{9}. In fact, while numerous theoretical works on the
subject of quantum chaos were carried out, their experimental realization
remains somewhat scarce. Moreover, there are still many questions to be
answered, yet, and there is obvious need for new systems showing chaos, and
new experiments.

In this paper we propose a new kind of physical systems showing dissipative
chaotic dynamics. These systems are modeled by a dissipative nonlinear
oscillator driven by two periodic forces of different frequencies. This
model was proposed to study quantum stochastic resonance in the authors'
previous paper \cite{10}, where it was shown, in particular, that the model
is available for experiments. It can be implemented at least for dynamics of
strongly interacting photons in optical cavity with $\chi ^{(3)}$ nonlinear
medium \cite{11}, and for cyclotron oscillations of a single electron in a
Penning trap \cite{12,13}. We would especially prefer to study the
transition to quantum chaos and its control, the role of dissipation and
quantum entanglement in chaotic dynamics, and adjacent questions of
characterizing quantum chaos. These investigations are complemented by
consideration of the information aspect of chaotic dynamics through the von
Neumann entropy.

In classical mechanics standard characterization of chaos might be given in
terms of the unpredictability of phase-space trajectories. However, the most
important characteristics of classical chaotic systems - exponential
divergence of trajectories, starting at arbitrarily close initial points in
phase-space - does not have quantum counterpart. The question of what the
quantum mechanical equivalent of chaos constitutes, has been posed. Many
criteria have been suggested to define chaos in quantum systems varying in
their emphasis and domain of application \cite{14}. As yet, there is no
universally accepted definition of quantum chaos. Our analysis of quantum
chaos is based mainly on the time-evolution of von Neumann entropy and
Wigner function.

The system under research is dissipative and therefore it has a mesoscopic
nature. Quite generally, chaos in classical conservative and dissipative
systems has completely different properties, e.g. strange attractors can
appear only in dissipative systems. Therefore, the system of our interest
might allow us to examine challenging problems of quantum dissipative chaos,
including the problem of quantum counterpart of a strange attractor. We
note, while the quantum dynamics of isolated or so called Hamiltonian
systems with chaotic classical counterparts has studied well, a very liitle
work has been done in looking at quantum chaos of dissipative nonlineat
systems. Among earlier studies of open quantum chaotic systems it should be
noted the papers \cite{15}. A new impetus to increasing the dissipative
quantum chaos area has recently given by study of decoherence and
quantum-classical correspondence problem of chaotic systems \cite{16,17}.

Below we use the traditional ensemble description of Markovian open systems,
based on the master equation. Then, this equation is presented in quantum
trajectories in the frames of quantum state diffusion approach (QSD) \cite
{18}. Recently, it was shown how quantum state diffusion can be used to
model dissipative chaotic systems on individual quantum trajectories \cite
{19,20}. In contrast with these papers, here we show how it is possible to
describe the quantum chaos using a statistical ensemble of trajectories,
which is actually realized in nature. Our results indicate, that though
properties of quantum chaotic dynamics do not appear on ensemble averaged
oscillatory excitation numbers, these properties are clearly obvious on the
entropy and Wigner function.

The outline of this paper is as follows: In the next section we describe the
system proposed, and give the analysis of its classical motion on the
Poincar\'{e} section in phase-space. In Sec.III we develop the quantum
description of the problem by numerically solving the master equation
through the QSD method. We present the averaged over quantum trajectories
results for the mean excitation number of nonlinear oscillator, and for both
the Wigner function and the von Neumann entropy. We summarize our results in
Section IV. \ 

\section{ Model of double driven oscillator: classical phase map}

In this section we give the theoretical description of the system. The
nonlinear oscillator driven by two periodic forces at frequencies $\omega
_{1}$ and$\ \omega _{2}$ and interacting with a reservoir is described by
the following Hamiltonian 
\begin{equation}
H=\hbar \omega _{0}a^{+}a+\hbar \chi (a^{+}a)^{2}+\hbar \left[ \left( \Omega
_{1}\exp (-i\omega _{1}t)+\Omega _{2}\exp (-i\omega _{2}t)\right) a^{+}+h.c.%
\right] +H_{loss},  \label{ham}
\end{equation}
\ where $a,a^{+}$ are boson annihilation and creation operators, $\omega
_{0} $ is an oscillatory frequency and $\chi $ is the strength of the
anharmonicity. The couplings with two driving forces are given by Rabi
frequencies $\Omega _{1}$ and $\Omega _{2}$. $H_{loss}=a\Gamma
^{+}+a^{+}\Gamma $ responsible for the linear losses of\ oscillatory states,
due to couplings with heat reservoir operators giving rise to the damping
rate $\gamma .$ The reduced density operator $\rho $ within the framework of
the rotating-wave approximation,\ in the interaction picture corresponding
to the transformation $\rho \rightarrow e^{-i\omega _{1}a^{+}at}\rho
e^{i\omega _{1}a^{+}at}$ is governed by the master equation

\begin{equation}
\frac{\partial \rho }{\partial t}=-\frac{i}{\hbar }\left[ H_{0}+H_{int,}\rho %
\right] +\sum_{i=1,2}\left( L_{i}\rho L_{i}^{+}-\frac{1}{2}%
L_{i}^{+}L_{i}\rho -\frac{1}{2}\rho L_{i}^{+}L_{i}\right) ,  \label{mastereq}
\end{equation}
where 
\begin{eqnarray}
H_{0} &=&\hbar \Delta a^{+}a,  \label{Volod} \\
H_{int} &=&\ \hbar \left[ \left( \Omega _{1}+\Omega _{2}\exp \left( -i\delta
t\right) \right) a^{+}+\left( \Omega _{1}^{\ast }+\Omega _{2}^{\ast }\exp
\left( i\delta t\right) \right) a\right] +\hbar \chi (a^{+}a)^{2}.  \nonumber
\end{eqnarray}
\ \ \ \ \ \ \ \ Here$\ \Delta =\omega _{0}-$ $\omega _{1}$\ is the detuning,
and $\delta =\omega _{2}-$ $\omega _{1}$ is the difference between driving
frequencies, which works as modulation frequency. $L_{i}$ are the Lindblad
operators:

\begin{equation}
L_{1}=\sqrt{\left( N+1\right) \gamma }a,\;L_{2}=\sqrt{N\gamma }a^{+},
\label{Lindblad}
\end{equation}
where $\gamma $ is the spontaneous decay rate of the dissipation process,
and $N$ denotes the mean number of quanta of a heat bath. Note, only the
case of a vacuum reservoir, $N=0$, considered below.

For $\Omega _{2}=0$ this equation describes the single driven, dissipative
anharmonic oscillator, which is a well-known and archetypal model in
nonlinear physics \cite{21,22}. In the semiclassical approach and in
steady-state this system exhibits bistability, which appears as the
hysteresis behavior of the mean oscillatory number, versus either the
detuning $\Delta $\ or the strength of driving $\Omega _{1}$\ \cite{21}.
However, the hysteresis in the quantum-mechanical treatment disappears on
ensemble-averaged mean oscillatory number $n(t)=\left\langle
a^{+}(t)a(t)\right\rangle ,$ and bistability, manifests itself on individual
quantum trajectories, as noise-induced transitions between two possible
metastable states \cite{23}, as well as on the statistics of oscillatory
numbers \cite{24}.

In the case of double driven oscillator, when two external forces are both
present, the corresponding Hamiltonian (\ref{Volod}) includes explicit
time-dependence, even in the rotating-wave approximation. It may therefore,
be expected that the system presented above, exhibits regions of regular and
chaotic motion depending from the parameters: $\chi ,$ $\Delta ,$ $\Omega
_{1},$ $\Omega _{2}$ and $\gamma .$ To illustrate the operational regimes of
the oscillator, first we pay attention to the classical description. After
making the usual approximations, the classical limit of Eq.(\ref{mastereq})
becomes

\begin{equation}
\frac{d}{dt}\alpha (t)=-\frac{1}{2}\gamma \alpha -i\left( \Delta +\chi
(1+2\left| \alpha \right| ^{2})\right) \alpha -i\left( \Omega _{1}+\Omega
_{2}\exp \left( -i\delta t\right) \right) ,  \label{clas}
\end{equation}
where $\alpha $\ is the dimensionless complex amplitude, corresponding to
the operator $a.$ We analyze the time-dependent solution of this equation in
the phase-space of dimensionless position and momentum $X=%
\mathop{\rm Re}%
\alpha $, $Y=%
\mathop{\rm Im}%
\alpha $. There are a number of ways to make such analysis.\ We adopt a
discrete surface or Poincar\'{e} section of this system. Let $X_{0},$ $Y_{0}$
be an arbitrary initial phase-space point of the system at the time $t_{0}.$
Then we define a constant phase map in the $(X,Y)$ plane by the sequence of
points $(X_{n},Y_{n})=\left( X(t_{n}),Y(t_{n})\right) ,$ changing the time
intervals by $t_{n}=t_{0}+\frac{2\pi }{\delta }n,$ $n=0,1,2,...$ . This
means, that for any $t=t_{n}$ the system is in one of the points of Poincar%
\'{e} section.\ Our analysis shows that for extended time scales exceeding
the damping rate, the asymptotic dynamics of the system is regular in the
limits of small and large values of modulation frequency, i.e. $\delta \ll
\gamma $ and $\delta \gg \gamma $, and also when one of the perturbation
forces is much greater than the other: $\Omega _{1}\ll \Omega _{2}$ or $%
\Omega _{2}\ll \Omega _{1}.$ The dynamics of the system is chaotic in the
range of parameters $\delta \gtrsim \gamma $ and\ $\Omega _{1}\simeq \Omega
_{2}$. Fig.1 shows the results of numerical calculations of the classical
maps, for the parameters chosen in the range of chaos. These figures clearly
indicate the classical strange attractors with fractal structure, which are
typical for a chaotic Poincar\'{e} section (we choose $t_{0}=0$ for all
cases). It is expected from these results, that the domain of phase-space
which includes an attractor, strongly depends on the parameters $\Omega _{1},
$ $\Omega _{2}$ and $\gamma $. In particular, the domain increases with the
increase of $\Omega _{1}$ or $\Omega _{2},$ and with the decrease of the
decay rate $\gamma $. Below we give the general consideration of this effect
considering equation (\ref{clas}) in the following integral form: 
\begin{equation}
\alpha (t)=e^{if(t)-\gamma t/2}\left[ -i\int\limits_{t_{0}}^{t}\left( \Omega
_{1}+\Omega _{2}e^{-i\delta t_{1}}\right) e^{if(t_{1})}e^{\gamma
t_{1}/2}dt_{1}+\alpha (t_{0})e^{\gamma t_{0}/2}\right] .  \label{integ}
\end{equation}
Here $\alpha _{0}=\alpha (t_{0})$ is an initial value at $t=t_{0},$ and $f(t)
$ is introduced as follows : 
\begin{equation}
f(t)=\int\limits_{t_{0}}^{t}\left( \Delta +\chi \left( 1+2\left| \alpha
(t_{1})\right| ^{2}\right) \right) dt_{1}.  \label{f(t)}
\end{equation}

It is easy to estimate the module of the complex amplitude $\left| \alpha
(t)\right| ,$\ taking into account that $f(t)$\ is a real function. For its
maximum on time interval value $\alpha _{\max }=\max_{t}\left| \alpha
(t)\right| $ we obtain 
\begin{equation}
\alpha _{\max }\leq \frac{\left( \left| \Omega _{1}\right| +\left| \Omega
_{2}\right| \right) }{\gamma }.  \label{amax}
\end{equation}

This formula determines the border of a classical map in dependence from
Rabi frequencies and the decay rate, and explains the above conclusion about
the size of the strange attractor in phase-space. It is interesting, that
this border size for the anharmonic oscillator is independent from the
parameter of nonlinearity $\chi $ and the detunings $\Delta $ and $\delta .$
It should be noted a peculiarity of the system proposed regarding the
strange attractors. As can be seen from the numerical results, the
attractors depicted on Fig. 1 (a) and Fig. 1 (c) for different parameters
have the same form in phase-space and differ from each other only on the
scales. It is obvious that this property is the consequence of a scaling
symmetry of the classical equation (\ref{clas}). Indeed, it easy to verify
that Eq.(\ref{clas}) remains invariable for the following scaling
transformation of the complex amplitude $\alpha \rightarrow \alpha ^{\prime
}=\lambda \alpha ,$ where $\lambda $ is a real positive dimensionless
coefficient, if the parameters $\chi ,$ $\Delta $, $\Omega _{1},$ $\Omega
_{2}$ correspondingly transformed as: $\chi \rightarrow \chi ^{\prime }=\chi
\diagup \lambda ^{2},$ $\Delta \rightarrow \Delta ^{\prime }=\Delta +\chi
\left( 1-1\diagup \lambda ^{2}\right) ,$ $\Omega _{1,2}\rightarrow \Omega
_{1,2}^{\prime }=\lambda \Omega _{1,2}.$\ We have illustrated such symmetry
on Figs. 1 (a),(c) where the attractors are presented for two sets of
parameters coupled by the scaling transformations. We note, that in the
quantum treatment the diffusion term in the master equation (\ref{mastereq})
affects the scaling symmetry roughly.\ In Sec.III, we will discuss this and
other properties of strange attractors common with the Wigner function, in
more detail.

\section{Quantum signatures of chaos: \ Entropy and Wigner function}

In this section we examine the problem of quantum dissipative chaos on the
basis of the von Neumann entropy and the Wigner function. The peculiarities
and advantages of such a consideration for interpreting quantum chaos in an
ensemble theory, are as follows:

The von Neumann entropy, which is defined through the reduced density
operator as 
\begin{equation}
S=-Tr\left( \rho \ln \rho \right) ,  \label{entropy}
\end{equation}
is a measure of dissipation and decoherence. The entropy for an isolated
quantum system, does not change under the time evolution. If the time
evolution of the system is perturbed through interaction with an
environment, the averaging over the perturbation typically leads to an
entropy increase. The von Neumann entropy also is a sensitive operational
measure of an entanglement, as well as the measure of the purity of quantum
states \cite{25}. If the system is in pure state, the entropy is precisely
zero. Thus, it is expected that the study of entropy (\ref{entropy}) for
doubly driven oscillator will allow to examine chaos in terms of quantum
entanglement.

For an extended time-evolution period the classically chaotic system at
moments $t_{n}=t_{0}+\frac{2\pi }{\delta }n$, $n=0,1,2,...$, fills the
Poincar\'{e} section in phase-space. There are a number of methods to treat
Poincar\'{e} section quantum mechanically. We note the Ref.\cite{26}, where
a method of quantization of classical dissipative maps was proposed. The
Poincar\'{e} section of one quantum trajectory was considered in \cite{19,20}%
. For our dynamical model we adopt the method of the Wigner function and
study the features of correspondence between the Wigner function, which is
taken at one of the moments $t=t_{n}=t_{0}+\frac{2\pi }{\delta }n,$ and
Poincar\'{e} section. It was obtained through a computer simulation of the
Wigner function.

We analyze the problem of dissipation on the basis of the QSD method, which
operates with stochastic states $\left| \Psi _{\xi }(t)\right\rangle ,$
describing the evolution along a quantum trajectory. The equation of motion
is: 
\begin{eqnarray}
\left| d\Psi _{\xi }\right\rangle &=&-\frac{i}{\hbar }\left(
H_{0}+H_{int}\right) \left| \Psi _{\xi }\right\rangle dt-  \label{QSD} \\
&&\frac{1}{2}\sum_{i}\left( L_{i}^{+}L_{i}-2\left\langle
L_{i}^{+}\right\rangle L_{i}+\left\langle L_{i}\right\rangle \left\langle
L_{i}^{+}\right\rangle \right) \left| \Psi _{\xi }\right\rangle
dt+\sum_{i}\left( L_{i}-\left\langle L_{i}\right\rangle \right) \left| \Psi
_{\xi }\right\rangle d\xi _{i}.  \nonumber
\end{eqnarray}
Here $\xi $ indicates the dependence on the stochastic process, the complex
Wiener variables $d\xi _{i}$ satisfy the fundamental correlation properties:

\begin{equation}
M\left( d\xi _{i}\right) =0,\;M\left( d\xi _{i}d\xi _{j}\right) =0,\;M\left(
d\xi _{i}d\xi _{j}^{\ast }\right) =\delta _{ij}dt,  \label{Expectation}
\end{equation}
and the expectation value equals $\left\langle L_{i}\right\rangle
=\left\langle \Psi _{\xi }\left| L_{i}\right| \Psi _{\xi }\right\rangle $.
According to this method the reduced density operator is calculated as the
ensemble mean 
\begin{equation}
\rho (t)=M\left( \left| \Psi _{\xi }\right\rangle \left\langle \Psi _{\xi
}\right| \right) =\lim_{m\rightarrow \infty }\frac{1}{m}\sum_{\xi
}^{m}\left| \Psi _{\xi }(t)\right\rangle \left\langle \Psi _{\xi }(t)\right|
\label{density}
\end{equation}
over the stochastic pure states $\left| \Psi _{\xi }(t)\right\rangle ,$
describing the evolution along a quantum trajectory.

Let us first qualitatively describe the most important physical processes,
which are responsible for the origin of chaotic dynamics in our QSD
numerical study. There are two ways to realize the controlling transition
from the regular to chaotic dynamics, by changing one of the parameters of
the system. One of these ways is to vary the strength $\Omega _{2}$ of the
second force in the range from $\Omega _{2}\ll \Omega _{1}$ to $\Omega
_{2}\gg \Omega _{1}$. In the limit $\Omega _{2}\ll \Omega _{1}$ the system
is reduced effectively to the model of single driven anharmonic oscillator,
which exhibits bistability for the definite range of parameters $\ \chi ,$ $%
\Delta ,$ $\Omega _{1},$ and $\gamma .$\ In this limit our analysis of the
time-dependent stochastic trajectories for expectation numbers $n_{\xi }(t)=$
$\left\langle \Psi _{\xi }\left| a^{+}a\right| \Psi _{\xi }\right\rangle $\
shows, that the system in the bistability range spends most of the time
close to one of the semiclassical solutions of Eq.(\ref{clas}), with quantum
interstate transitions, occurring at random intervals. With increasing the
amplitude $\Omega _{2}$ for the other parameters, chosen to lead bistability
in terms of semiclassical solution, the stimulated processes, i.e. dynamical
interstate transitions, become sufficient. Their contribution at $\Omega
_{2}\simeq \Omega _{1}$\ leads to the emergence of a chaotic regime on
quantum trajectories. In the limit $\Omega _{2}\gg \Omega _{1}$ the regular
dynamics is restored. This limit is equivalent to the case of $\Omega
_{1}\gg \Omega _{2},$ because we can choose such an interaction picture,
that time dependent exponent in (\ref{Volod}) will appear near $\Omega _{1}.$
In fact, using the transformation $\rho \rightarrow e^{-i\omega
_{2}a^{+}at}\rho e^{i\omega _{2}a^{+}at}$ of the reduced density operator,
we arrive to the master equation (\ref{mastereq}) with the Hamiltonian 
\begin{eqnarray}
H_{0}^{^{\prime }} &=&\hbar (\Delta -\delta )a^{+}a,  \label{ham`} \\
H_{int}^{^{\prime }} &=&\ \hbar \left[ \left( \Omega _{2}+\Omega _{1}\exp
\left( i\delta t\right) \right) a^{+}+\left( \Omega _{2}^{\ast }+\Omega
_{1}^{\ast }\exp \left( -i\delta t\right) \right) a\right] +\hbar \chi
(a^{+}a)^{2}.  \nonumber
\end{eqnarray}
It is obtained from (\ref{Volod}) by the replacements $\omega
_{1(2)}\rightarrow \omega _{2(1)},$ $\Omega _{1(2)}\rightarrow $\ $\Omega
_{2(1)},$ $\Delta \rightarrow \ \Delta -\delta $. Nevertheless, these
marginal cases differ in details, as explained below.

In an other scenario of transition to chaos the modulation frequency $\delta 
$ is varied, with other unchanged parameters. In the range of small
frequencies $\delta \ll \gamma $ the modulation of the system is adiabatic.
So, in the range of bistability the system oscillates between the two
possible metastable states. With increasing frequency, at $\delta \gtrsim
\gamma $, a strong entanglement of these states occurs, and the system comes
to chaos. It should be noted that, as we will show below, the transition
from the regular to chaotic dynamics in the classical system, is marked in
the quantum system by an increase of the von Neumann entropy, as well as by
a strong transformation of the Wigner functions. For the case $\delta \gg
\gamma ,$ the dynamics of the system becomes regular again.

We give the results of QSD analysis in the regimes of strong anharmonicity,
considering the parameters from $\chi /\gamma =0.7$ to $\chi /\gamma =0.1$.
The former case is strongly quantum mechanical, since the maximum mean
number of oscillatory excitations is about 10, while the case, $\chi /\gamma
=0.1$, corresponds to a quasiclassical regime, when the maximum in time
oscillatory number is about 130. The truncated basis of Fock number states
of harmonic oscillator is used for the expansion of state vectors $\left|
\Psi _{\xi }(t)\right\rangle ,$ and an initial vacuum state is chosen.

\subsection{Time-evolution of the mean oscillatory numbers}

The mean excitation number of double driven anharmonic oscillator versus
time interval is of particular interest in this paper. Figures 2(a),(b)
depict the ensemble averaged mean oscillatory numbers $\overline{n}%
=\left\langle a^{+}a\right\rangle $ for both cases of regular and chaotic
dynamics, which are realized in the classical limit. The parameters for
Fig.2(b) are chosen the same as for Fig.1 (c). The classical oscillatory
number $\overline{n}=\left| \alpha \right| ^{2}$ derived from equation (\ref
{clas}) for the same parameters as the above chaotic regime, is illustrated
in Fig.2(c). From these figures it is evident that while the classical
result (Fig.2(c)) shows usual chaotic behavior, its quantum ensemble
counterpart (Fig.2(b)) has clear regular behavior. Thus, the chaotic
behavior in the classical model, transforms into the periodic dynamics in
the quantum treatment which involves ensemble averaging. These results
indicate, that quantum dissipative chaotic dynamics is not evident on the
mean oscillatory number. Now we will show how quantum chaos emerges on both
the Wigner function and the von Neumann entropy.

\subsection{Chaos on Wigner functions and quantum interference effect}

We apply the QSD to determine Wigner functions for the quantum states of
double driven anharmonic oscillator during time evolution. For this we use
the well-known expression for the Wigner function in terms of matrix
elements $\rho _{nm}=\left\langle n\left| \rho \right| m\right\rangle $ of
density operator in the Fock state representation 
\begin{equation}
W(r,\theta )=\sum_{m,n}\rho _{nm}W_{mn}(r,\theta ),  \label{wig}
\end{equation}
where $(r,\theta )$ are the polar coordinates in the complex phase-space
plane $X=r\cos \theta $, $Y=r\sin \theta ,$ and the coefficients $%
W_{mn}(r,\theta )$ are Fourier transform of matrix elements of the Wigner
characterization function\thinspace \cite{27} 
\begin{equation}
W_{mn}(r,\theta )=%
{\frac{2}{\pi }(-1)^{n}\sqrt{\frac{n!}{m!}}e^{i(m-n)\theta }(2r)^{m-n}e^{-2r^{2}}L_{n}^{m-n}(4r^{2}),\;m\geq n \atopwithdelims\{\} \frac{2}{\pi }(-1)^{m}\sqrt{\frac{m!}{n!}}e^{i(m-n)\theta }(2r)^{n-m}e^{-2r^{2}}L_{m}^{n-m}(4r^{2}),\;n\geq m}%
,  \label{wcoef}
\end{equation}
where $L_{p}^{q}$ are Laguerre polynomials. In our calculation we assume,
that the oscillator is initially prepared in a vacuum state, and regimes of
strong anharmonicity $\chi /\gamma =(0.7\div 0.1)$\ are realized.

In Fig.3 we demonstrate moving of our system from the regular to chaotic
dynamics by plotting the Wigner function for three values of $\Omega _{2}$: $%
\Omega _{2}/\gamma =1$ (a)$,$ $\Omega _{2}/\gamma =\Omega _{1}/\gamma =10.2$
(b)$,$ $\Omega _{2}/\gamma =20$ (c), in a fixed moment of time. The values
of $\Delta /\gamma ,$ $\chi /\gamma ,$ and $\Omega _{1}/\gamma $ are chosen
to lead to bistability in the model of single driven oscillator $(\Omega
_{2}=0)$.

We can see that for the case of a weak second force Fig.3(a) the Wigner
function has two humps, corresponding to the lower and upper level of
excitation of anharmonic oscillator in the bistability region. The hump
centered close to $X=Y=0$ describes the approximately coherent lower state,
while the other hump shows, that the upper state is squeezed. The effect of
squeezing is displayed as the squeezing of Gaussian. This result represents
the known property of the Wigner function for the single driven anharmonic
oscillator \cite{28}. The graphs in Fig.3 are given at an arbitrary time,
exceeding the damping time. As calculations show, for the next time
intervals during the period of modulation $t=2\pi /\delta ,$ the hump
corresponding to the upper level rotates around the central peak. When we
increase the strength of the second force, the classical system reaches to a
chaotic dynamics. The Wigner function for the chaotic dynamics is depicted
in Fig.3(b). Further increasing $\Omega _{2},$ the system returns to the
regular dynamics. The corresponding Wigner function at an arbitrary time
exceeding the transient time is presented in Fig.3(c). It contains only one
hump, rotating around the centre of the phase-space within the period. As
mentioned above, the limit $\Omega _{2}\gg \Omega _{1}$ is physically
equivalent to the opposite case $\Omega _{1}\gg \Omega _{2}$, when the
system in each moment of time is close to the model of single driven
nonlinear oscillator. The difference is that the Rabi frequency $\Omega _{2}$%
\ in this case (Fig.3(c)) is taken outside the bistability range, where the
system is in the upper level. So, the Wigner function has approximately the
same form, as the Wigner function for single driven anharmonic oscillator,
in the monostable above threshold regime of operation, when the system is
excited.

As we see, the Wigner function reflecting chaotic dynamics (Fig.3(b)), has a
complicated structure. Nevertheless, it is easy to observe that its contour
plots in $(X,Y)$ plane are generally similar to the corresponding classical
Poincar\'{e} section. Now we will consider this problem in more detail,
comparing the results summarized in Fig.1 with the numerical calculations of
Wigner functions, for the same sets of parameters as for the classical maps.

We present our results in Fig.4. It can be seen in Fig.4.(a) that for the
deep quantum regime ($\chi /\gamma =0.7,$ $\Delta /\gamma =-15,$ $\delta
/\gamma =5$), the contour plot of the Wigner function is smooth and
concentrated approximately around the attractor (Fig.1(a)). Nevertheless,
the different branches of the attractor are hardly resolved in Fig.4.(a). It
can also be seen, that in this deep quantum regime, an\ enlargement of the
Wigner function occurs in contrast to the Poincar\'{e} section.

Taking a smaller $\chi /\gamma $, the contour plot of the Wigner function
approaches the classical Poincar\'{e} section. This can be seen in
Figs.4(b),(c). For the last case the correspondence is maximal, and some
details of the attractor (Fig.1(c)) are resolved much better in Fig.4(c).
This analysis allows us also to note that the scaling symmetry of strange
attractors shown on Figs. 1\ (a), (c) dissapears for corresponding contour
plots of Wigner functions [Figs. 4 (a), (c)] in the quantum treatment of
dissipative chaos.

It should be specified that for all contour plots in Fig.4, the
corresponding Wigner functions have regions of negative values. These
results are related to chaotic regimes. Obviously, this fact reflects a
quantum interference in the chaotic state of the system considered. It
follows that the ranges of negative values of the Wigner function increase
with the increasing parameter $\chi /\gamma ,$ when the system moves to a
deep quantum regime. Nevertheless, the ranges of negative values of the
Wigner function are also observed for the comparatively small parameter $%
\chi /\gamma $, where an operation regime close to the semiclassical is
realized. It is illustrated in Fig.5 for parameter $\chi /\gamma =0.1$,
where the mean excitation number equals 130.\ 

Since our model is dynamical, we can also consider the correspondence
between the Wigner function and the Poincar\'{e} section, which are taken at 
$t_{n}=t_{0}+\frac{2\pi n}{\delta }$, for arbitrary $t_{0}$. Despite
different forms, which the Wigner function and the Poincar\'{e} section have
acquired, the correspondence features are the same.

\subsection{The emergence of chaos on the von Neumann entropy}

As shown in Section IIIA, the quantum chaotic dynamics is not displayed
clearly on ensemble averaged oscillatory excitation number. The purpose of
this subsection is merely to demonstrate how chaos is seen on the von
Neumann entropy, which is one of the significant characteristics of a
quantum ensemble. We also clarify other important questions in relation with
quantum chaos and entanglement in the system. These analysis complete the
above studies of quantum dissipative chaos in the Wigner function.

Two ways of producing chaos in a controlled manner will be considered by
monitoring the system through varying either the strength of the driving
force, or the difference frequency $\delta $. We calculate the evolution of
the entropy by formula (\ref{entropy}), using the results for reduced
density matrix expressed through an ensemble of trajectories. The
calculations are performed by diagonalization of the matrix $\rho _{nm}$ in
the truncated Fock states basis.

First, (Fig.6) we demonstrate the transition of the system from regular
dynamics to chaos by plotting the von Neumann entropy for three values of
strength of the driving field:\ $\Omega _{2}/\gamma =1$ (a)$,$ $\Omega
_{2}/\gamma =\Omega _{1}/\gamma =10.2$ (b)$,$ $\Omega _{2}/\gamma =20$ (c).
The same parameters are chosen as in Fig.3 for the Wigner function. In
Fig.6(a) and Fig.6(c) we plot the entropy evolution for the regions with
classically regular behavior, while the Fig.6(b) shows entropy production
for the chaotic motion. The essential difference between the behavior of the
entropy for regular and chaotic dynamics is clearly displayed in these
figures. The common feature is, that for times exceeding the time scale of
transient dynamics, the entropy production in the chaotic regime dominates
over the entropy production of regular dynamics. These results are in good
qualitative agreement with the above results on the Wigner function, and
reflect the dependence of entropy production on both the quantum
entanglement, and formation of states in the system. Naturally, entropy
production is stipulated by the entangling interaction between the
anharmonic oscillator and its environment on the one hand, and is determined
by the structure of mixed states of nonlinear oscillator, on the other hand.
As a result, the maximal value of the entropy (Fig.6(b)) is realized for
chaotic dynamics with a large number of mixed states as depicted in
Fig.3(b), while its minimum takes place for the regular dynamics with one
hump Wigner function (Fig.3(c)). Also, we note, that the oscillations of the
entropies evident in Fig.5, have the frequency $\delta .$\ The simulations
also show the definite difference between transient times of regular and
chaotic dynamics. There is some ambiguity in the definition of the
transition time because of the oscillatory nature of curves. Nevertheless,
it is clearly evident, that the transient time of entropy evolution for the
regular case exceeds one for the chaotic dynamics.

In addition to providing criteria to characterize chaos, we have also
studied the behavior of the entropy versus controlling parameters $\Omega
_{2}$ and $\delta .$ The results of numerical calculations at a definite
time moment exceeding the transient time, are presented in Fig.7(a), where
minimum values of entropy for different $\Omega _{2}$ are presented. Let us
compare these results with the emergence of chaos in the classical limit. It
is easy to check, that for the classical case and for parameters shown in
Fig.6 chaos appears at the critical point of $\Omega _{2}=\Omega _{cr}\simeq
8.195,$ and disappears at $\Omega _{2}\simeq 12.745.$ It is a well known
property of classical chaos, that it appears suddenly. As it is seen in
Fig.7(a), the quantum chaos appears smoothly: the entropy increases, as the
value of $\Omega _{2}$ approaches the critical value $\Omega _{cr}.$ Another
way to probe chaos is the varying of detuning $\delta $. In Fig.7(b) the
behavior of the entropy versus modulation frequency $\delta $ are displayed.

\section{Conclusion}

We have presented novel type of a time-dependent systems showing chaotic
dynamics and intrinsically quantum properties. These systems are modelled by
a dissipative anharmonic oscillator driven by two forces of different
frequencies. We emphasize that this model is different from the ones like
single driven nonlinear oscillator, where a pulsed pump field could be used
and might be proposed as a possible experimental test of quantum chaos in
area of quantum optics with continuous cw laser. The proposed model seems
experimentally feasible with state-of the-art equipment and can be realized
at least in two experimental schemes. So, the nonlinear behavior of
single-mode field in a medium with a third-order nonlinearity may provide
simple realization of the dynamics of driven anharmonic oscillator. In fact,
a single-mode field is well described in terms of anharmonic oscillator, and
the nonlinear media could be an optical fiber or a $\chi ^{(3)}$ crystal,
placed in a cavity. In the later case the anharmonicity of mode dynamics
comes from the self-phase modulation due to the photon-photon interaction in 
$\chi ^{(3)}$-medium, and dissipative effects arise from the leakage of
photons through the cavity mirrors. Such a system under two driving fields
is described by the Hamiltonian (\ref{ham}) \ with $a,$ $a^{+}$ being the
operators of a single cavity mode. Cyclotron oscillations of a
single-electron in a Penning trap with a magnetic field, is another
realization of the quantum anharmonic oscillator. Its anharmonicity comes
from nonlinear relativistic correction to an electron motion, while
dissipative effects arise from the spontaneous emission of synchrotron
radiation \cite{12}. The trapped electron driven by a single coherent field,
has been experimentally realized and studied in Refs. \cite{13}. The model
we present here corresponds to the one-electron cyclotron oscillator in two
coherent fields at different frequencies. The corresponding Hamiltonian is
given by (\ref{ham}), where operators $a,$ $a^{+}$ describe the cyclotron
quantized motion at the cyclotron frequency. The values of the parameter $%
\chi /\gamma $ used in our calculations, have been achieved experimentally
for both above mentioned physical systems.

The dynamics of double driven anharmonic oscillator exhibits a rich
phase-space structure, including the regimes of regular and chaotic motion,
with the two Rabi frequencies $\Omega _{1}$ and $\Omega _{2}$, and the
difference $\delta $\ between driving frequencies being the control
parameters. We suppose that the adequate way of investigating the quantum
chaos is not only the investigation of the behavior of an individual
realization of trajectories, as suggested by several authors, but also
studying the dynamics of statistical ensemble of quantum trajectories, which
is naturally realized in experiments. For realization of this program of
studies the quantum state diffusion simulation method, based on the master
equation of Lindblad form is used. We also conclude, that the distinction
between regular and chaotic dynamics can be most easily understood by
studying the dynamics of essentially quantum properties, as the von Neumann
entropy and the Wigner function are only two examples. In fact, our
numerical analysis has shown that the quantum dynamical manifestation of
chaotic behavior, does not appear on ensemble averaged oscillatory
excitation numbers, but is clearly seen on the entropy and probability
distributions. The connection between quantum and classical treatments of
chaos was realized by means of comparison between strange attractors on the
classical Poincar\'{e} section, and the contour plots of the Wigner
functions. We have demonstrated that for small values of the ratio $\chi
\diagup \gamma $ the contour plots of Wigner functions are relatively close
to the strange attractors. Indeed, as we have shown on Fig. 1 (c) and Fig. 4
(c) some details of the attractor are resolved on the contour plots. Such
likeness of quantum and classical distribution vanishes in the deep quantum
regime [see Fig. 1 (a), and Fig. 4 (b) for $\chi \diagup \gamma =0.7$]. An
important point to emphasize is that the scaling symmetry of strange
attractors in the model studied here violates in the quantum treatment of
chaos. The drastic difference between the time-dependent behavior of the von
Neumann entropy for regular and chaotic regimes is clearly displayed in
Figs. 6. The von Neumann entropy indicates the connection between chaos and
entanglement, and has also been used to study characteristic time scales of
the emergence of chaos. For short time intervals $t<0.4\gamma ^{-1}$ the
entropies are linearly increasing function of time for both the regular and
chaotic regimes. For times exceeding the time scale of transient dynamics
this behavior transforms to the periodic one, however the entropy for the
chaotic regime dominates over the entropy for the regular dynamics. It has
also shown that the transient time for the chaotic regime is smaller than in
the regular regime.

In our analysis we have not investigated all possible quantum effects of
chaotic dynamics. In particular, we have specified that the Wigner function
for chaotic regime has regions of negative values even for relatively high
values of $\chi \diagup \gamma $ ($\chi \diagup \gamma =0.1$ on Fig. 5,
where the mean oscillatory number $n=130$). This fact reflects the quantum
interference effect in the chaotic dissipative dynamics. However, we have
not analyzed the correlation between the emergence of chaos and quantum
interference, which is an interesting albeit complicated option for the
future.

{\bf ACKNOWLEDGMENTS}

This work is partially supported by INTAS Grant No. 97-1672, and by Grant
No.00375 awarded by the Armenian Science Foundation.

\bigskip

\begin{center}
{\bf FIGURE CAPTIONS}
\end{center}

Fig.1. Poincar\'{e} section (approximately 20000 points)\ for the
dimensionless classical complex amplitude of double driven anharmonic
oscillator, plotted at times of constant phase $t_{n}\delta =$ $2\pi n$. The
dimensionless parameters are in the region of chaos:

(a) $\chi /\gamma =0.7,$ $\Delta /\gamma =-15,$ $\Omega _{1}/\gamma =\Omega
_{2}/\gamma =10.2,$ $\delta /\gamma =5;$

(b) $\chi /\gamma =0.5,$ $\Delta /\gamma =-25,$ $\Omega _{1}/\gamma =\Omega
_{2}/\gamma =25,$ $\delta /\gamma =15;$

(c) $\chi /\gamma =0.1,$ $\Delta /\gamma =-15,$ $\Omega _{1}/\gamma =\Omega
_{2}/\gamma =27,$ $\delta /\gamma =5.$

\bigskip

Fig.2. (a), (b) Ensemble averaged (over 1000 trajectories)\ mean oscillatory
numbers $\left\langle a^{+}a\right\rangle $\ versus dimensionless scaled
time $\gamma \tau $ for both regular (a) and chaotic (b) dynamics for
parameters: (a) $\chi /\gamma =0.1,$ $\Delta /\gamma =-15,$ $\Omega
_{1}/\gamma =\Omega _{2}/\gamma =27,$ $\delta /\gamma =50;$ (b) $\chi
/\gamma =0.1,$ $\Delta /\gamma =-15,$ $\Omega _{1}/\gamma =\Omega
_{2}/\gamma =27,$ $\delta /\gamma =5.$

(c) The classical oscillatory mean number $\left| \alpha \right| ^{2}$ for
chaotic dynamics. The parameters are the same as for (b).

\bigskip

Fig.3. Transition from regular to chaotic dynamics on the Wigner functions
for three values of $\Omega _{2}/\gamma :$ $\ \Omega _{2}/\gamma =1$ (a); $%
\Omega _{2}/\gamma =10.2$ (b); $\Omega _{2}/\gamma =20$ (c). The parameters
are: $\chi /\gamma =0.7,$ $\Delta /\gamma =-15,$ $\Omega _{1}/\gamma =\Omega
_{2}/\gamma =10.2,$ $\delta /\gamma =5.$ The averaging is over 2000
trajectories.

\bigskip

Fig.4. Contour plots of Wigner functions corresponding to chaos. The
parameters for the cases of (a),(b) and (c) are the same as in Figs.1(a),(b)
and (c) respectively. The averaging is over 2000 trajectories.

\bigskip

Fig.5.Illustration of quantum interference effect on the Wigner function for
the parameters: $\chi /\gamma =0.1,$ $\Delta /\gamma =-15,$ $\Omega
_{1}/\gamma =\Omega _{2}/\gamma =27,$ $\delta /\gamma =5.$ The averaging is
over 1000 trajectories.

\bigskip

Fig.6. Transtition to chaos on the von Neumann entropy: (a) and (c) -
regular dynamics, (b) - for chaotic dynamic. The averaging is over 2000
trajectories.

\bigskip

Fig.7. Behavior of the minimal (during the period)\ values of von Neumann
entropies versus controlling parameters.

(a). Dependence of entropy from $\Omega _{2}/\gamma $ for the parameters: $%
\chi /\gamma =0.7,$ $\Delta /\gamma =-15,$ $\Omega _{1}/\gamma =10.2,$ $%
\delta /\gamma =5.$

(b). Dependence of entropy from $\delta /\gamma $ for the parameters: $\chi
/\gamma =0.7,$ $\Delta /\gamma =-15,$ $\Omega _{1}/\gamma =\Omega
_{2}/\gamma =10.2.$ The averaging is over 2000 trajectories.

\bigskip

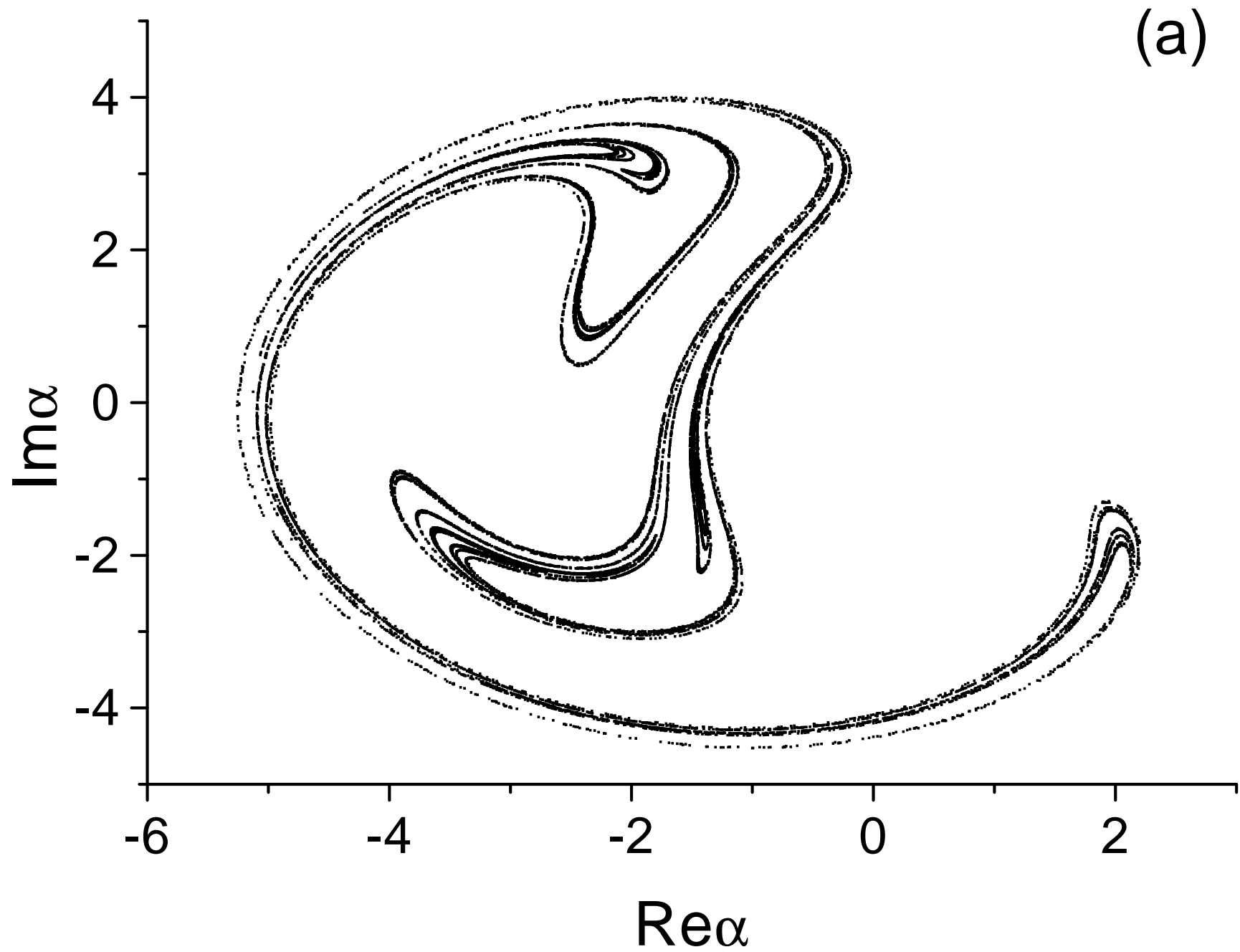
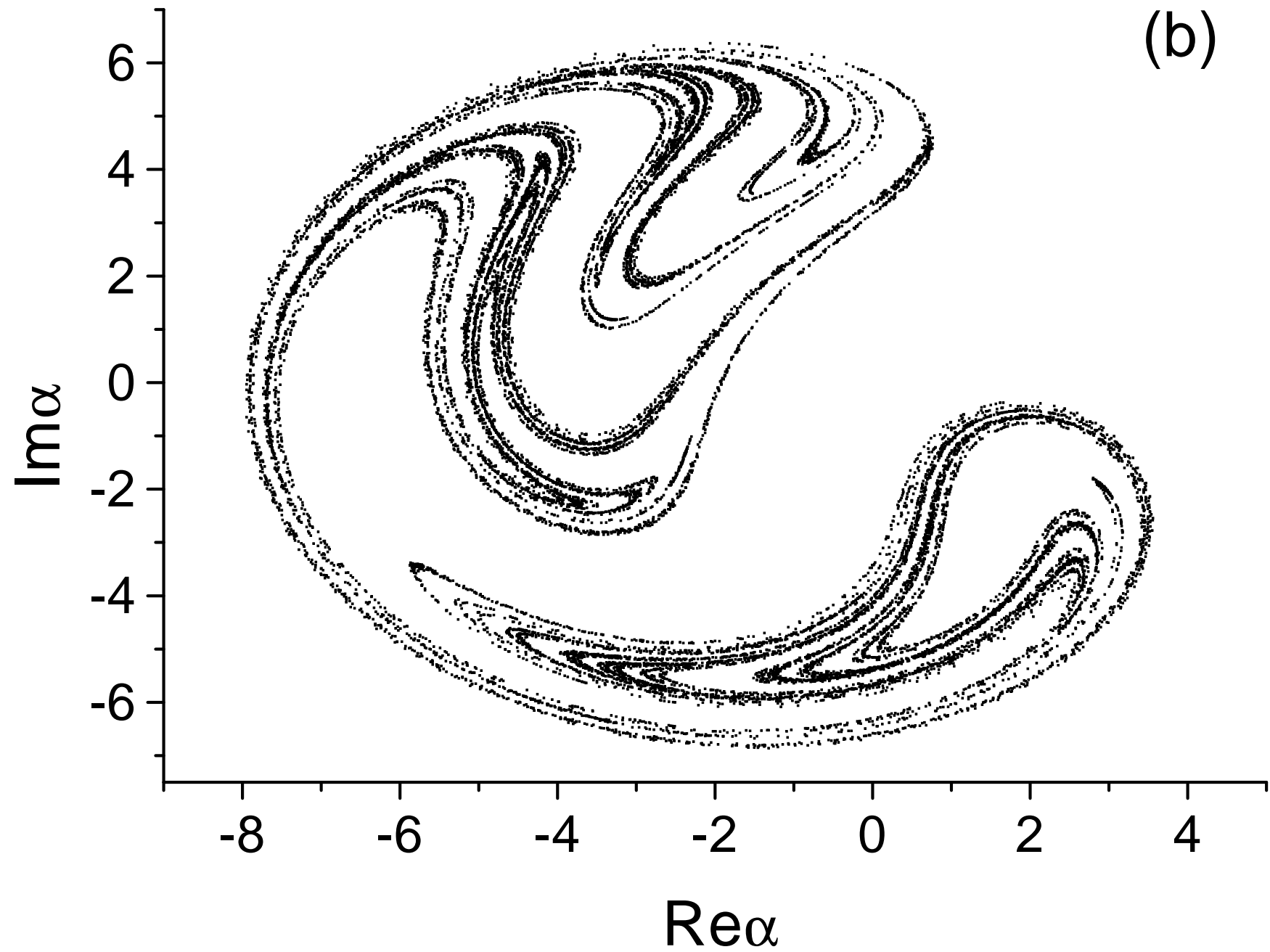
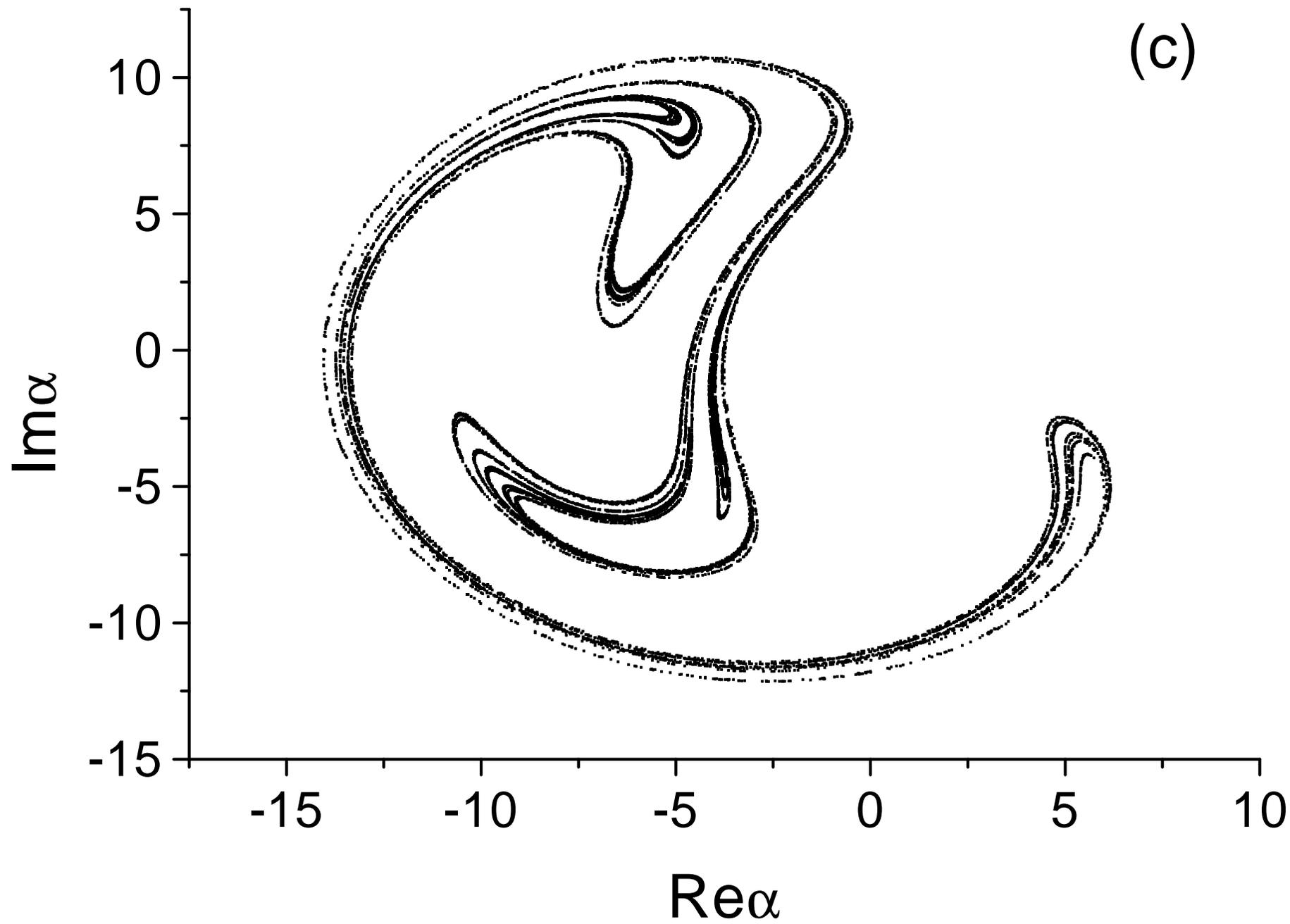
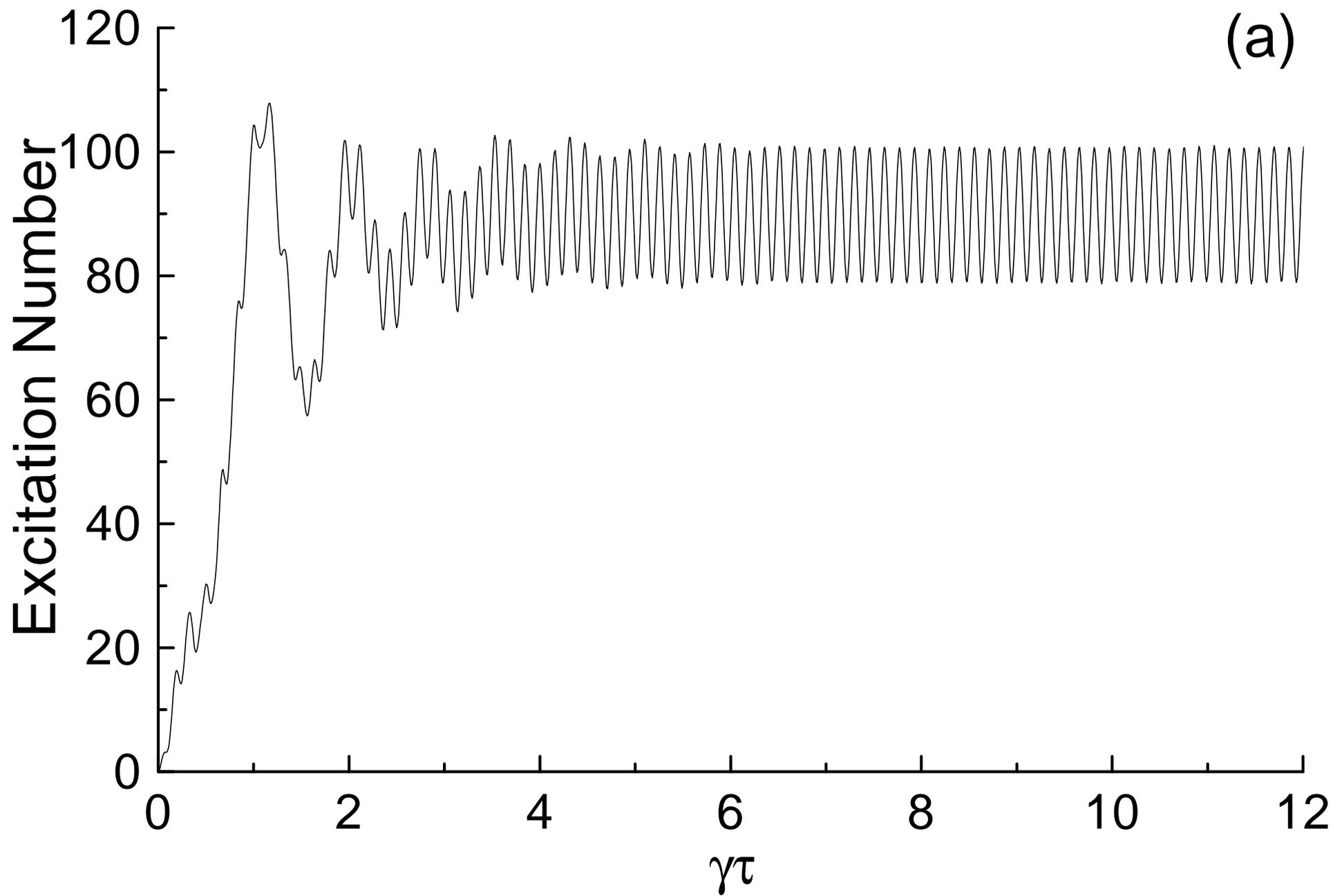
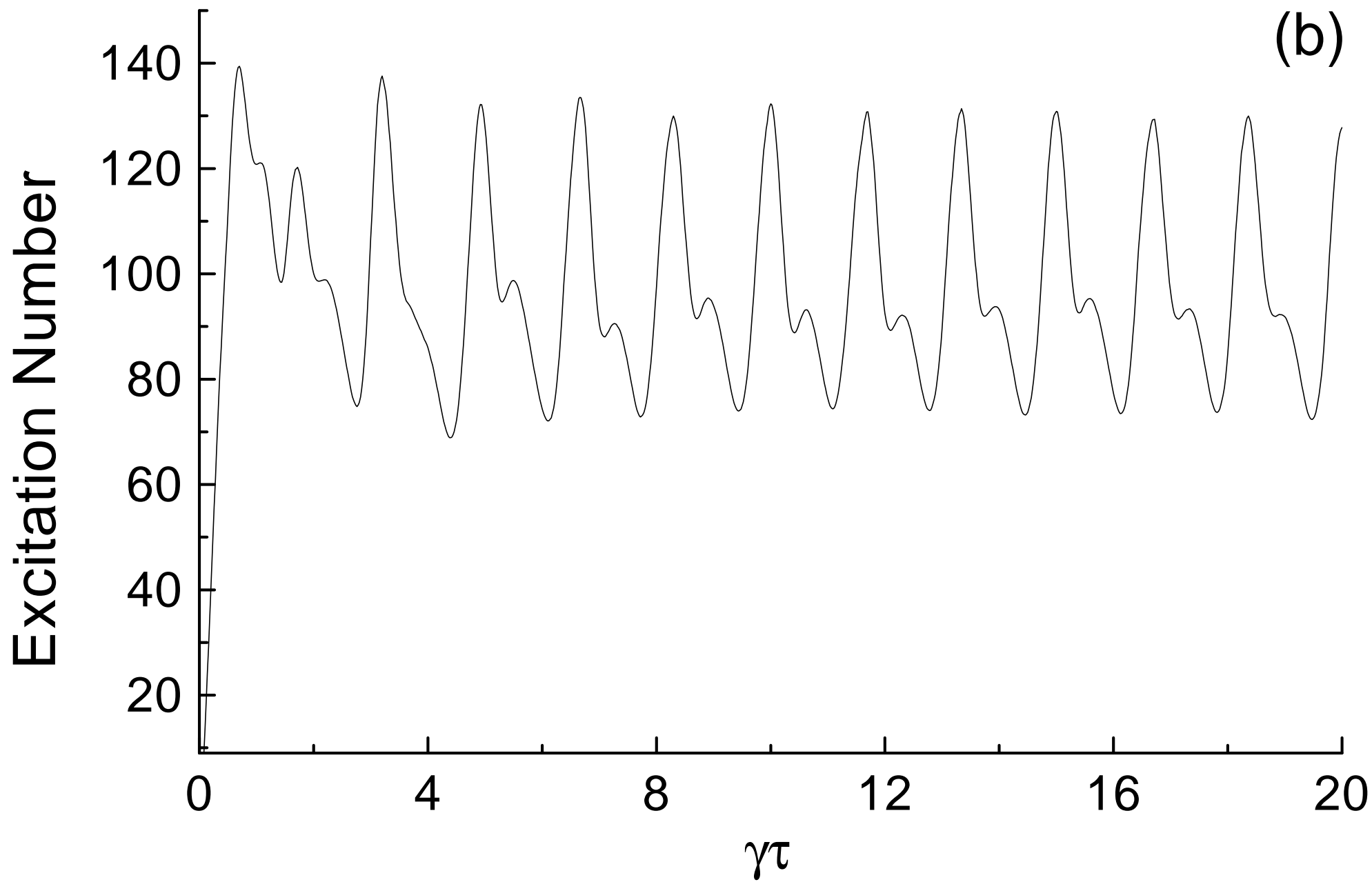
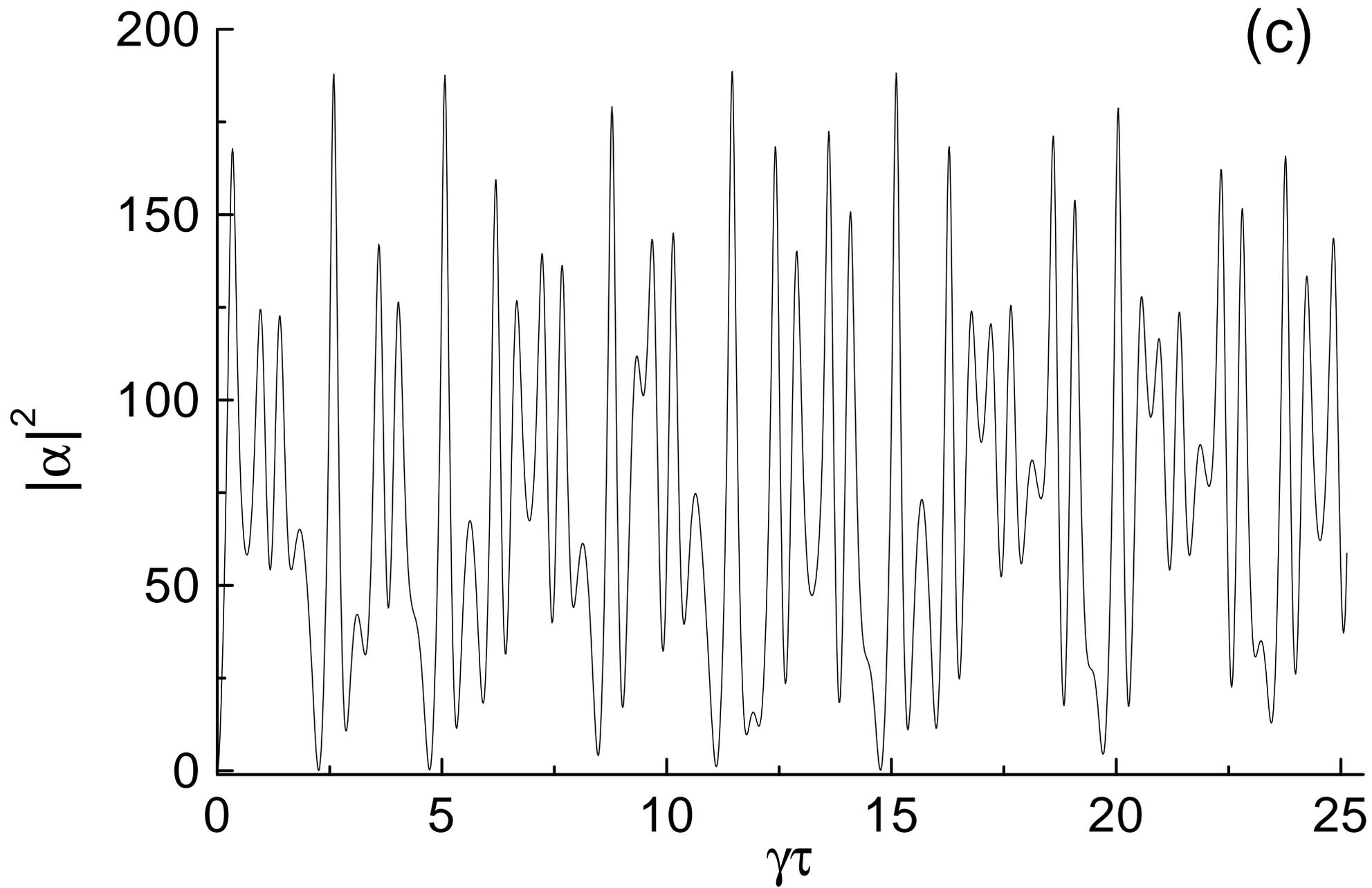
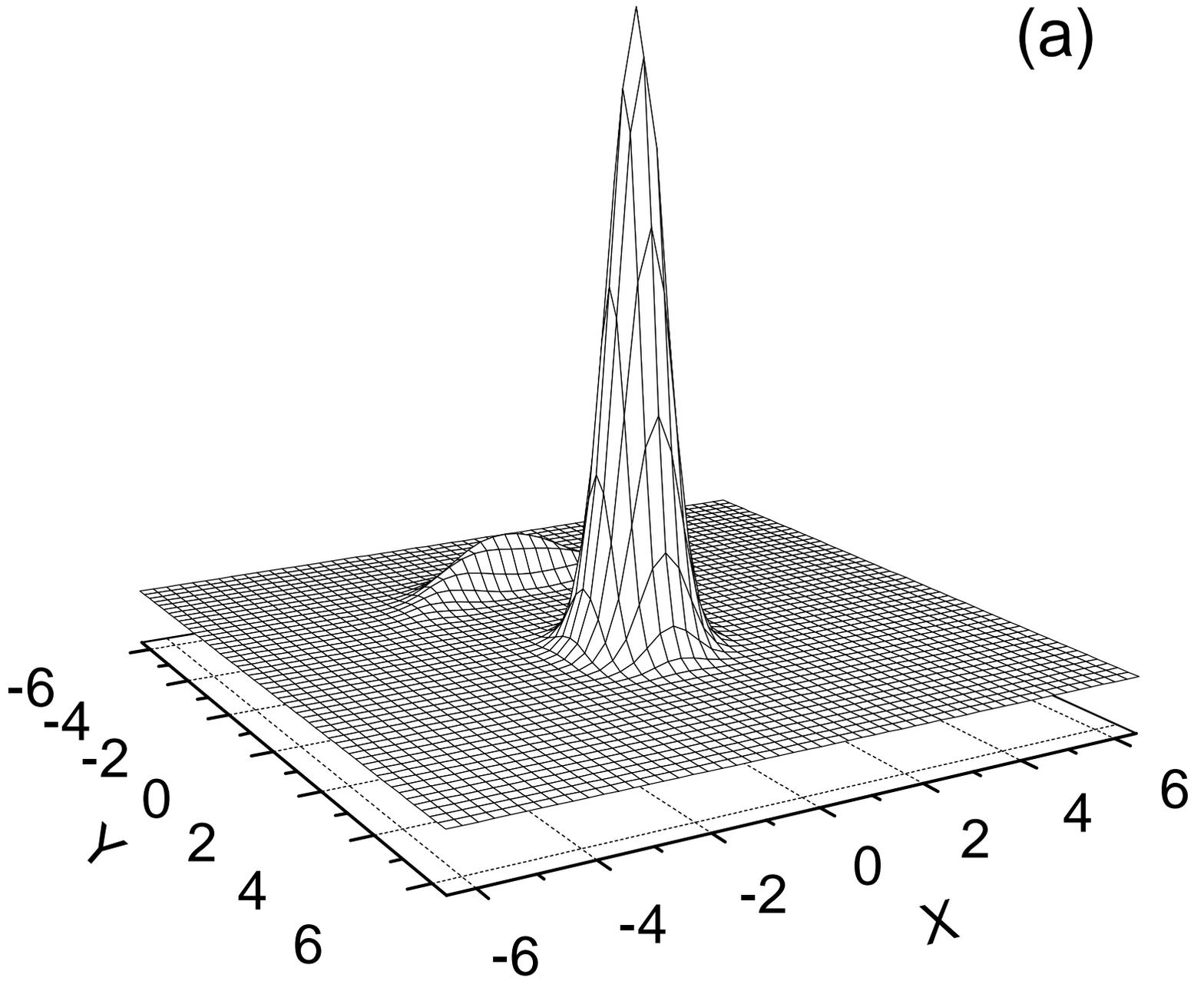
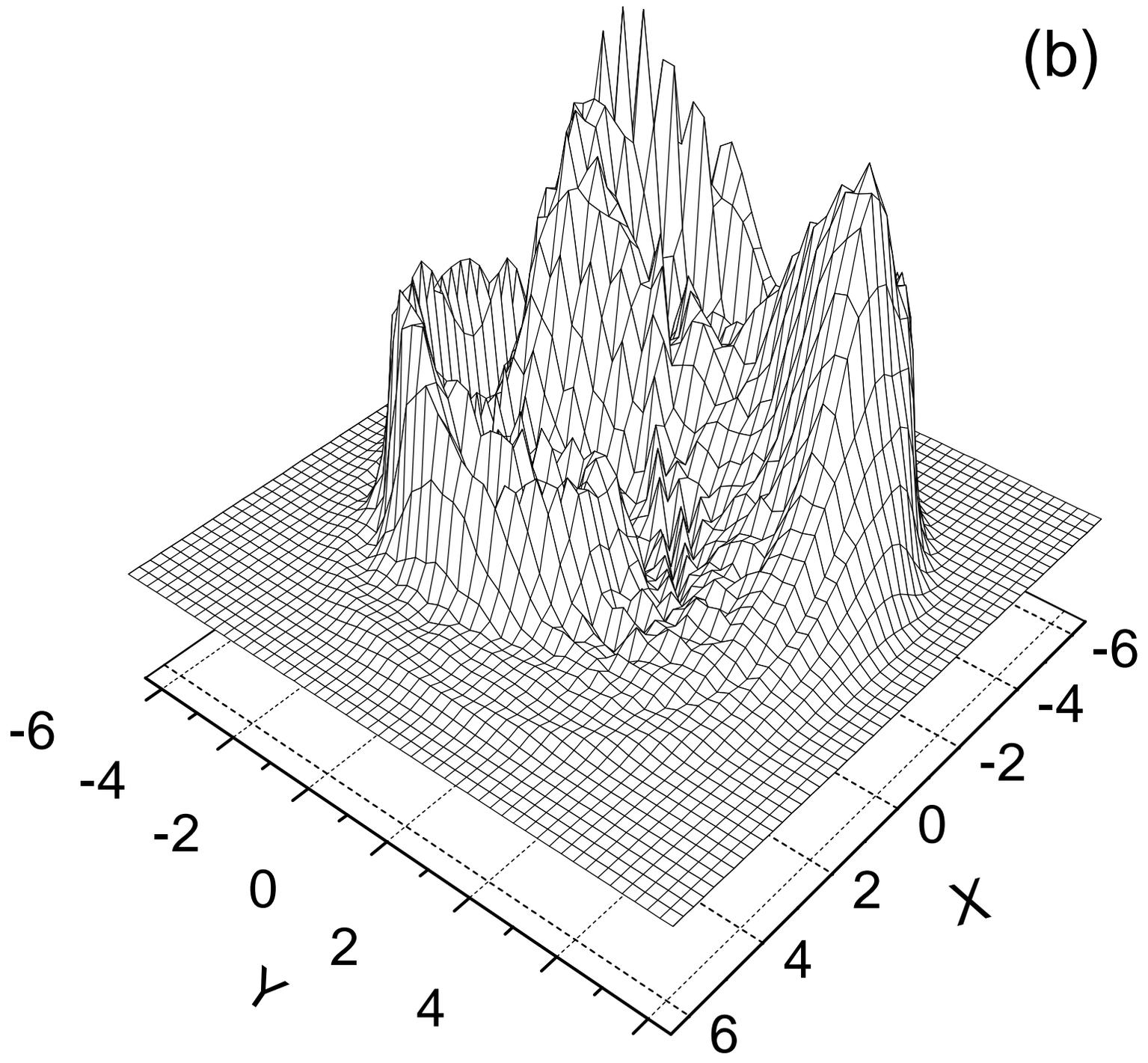
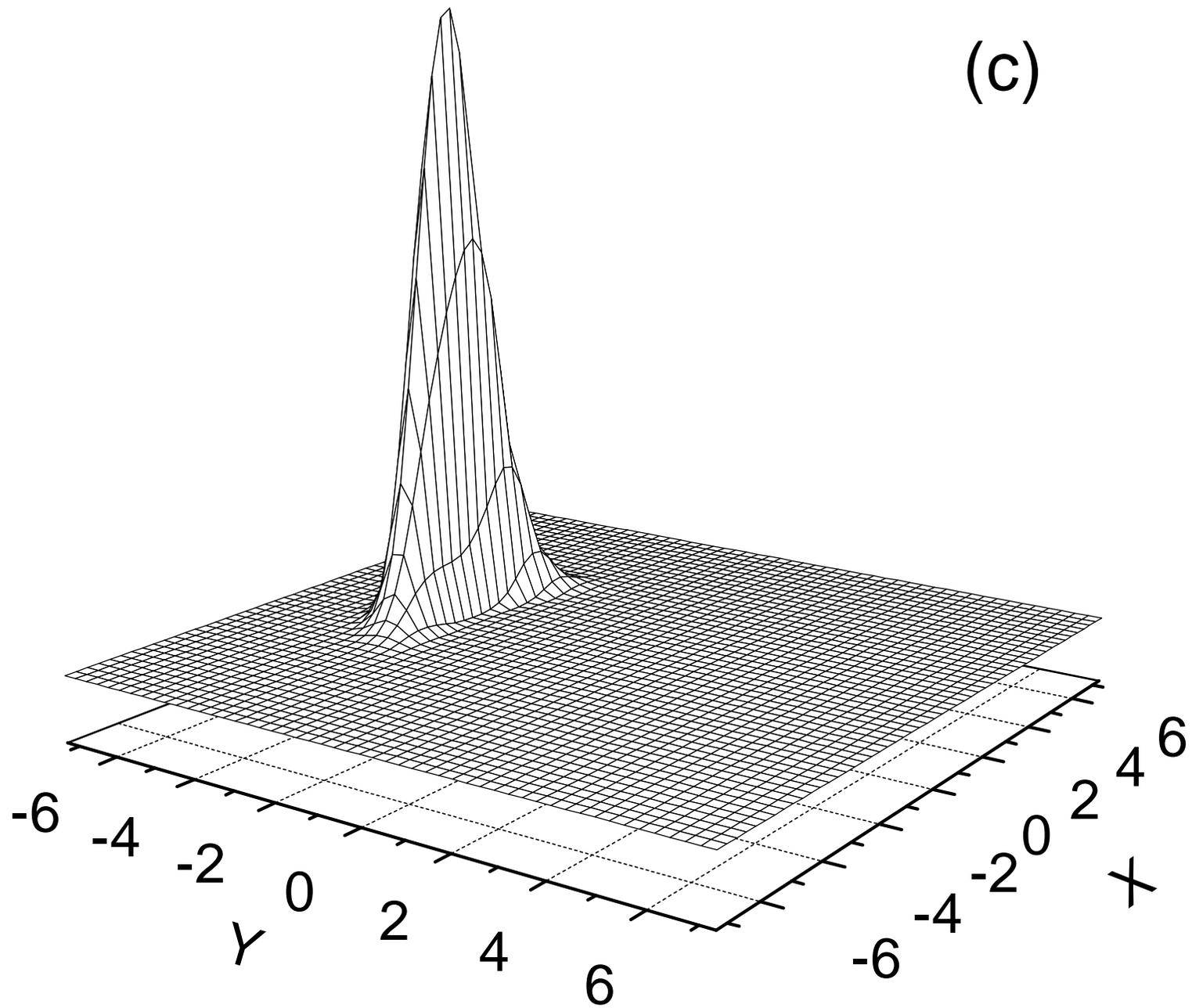
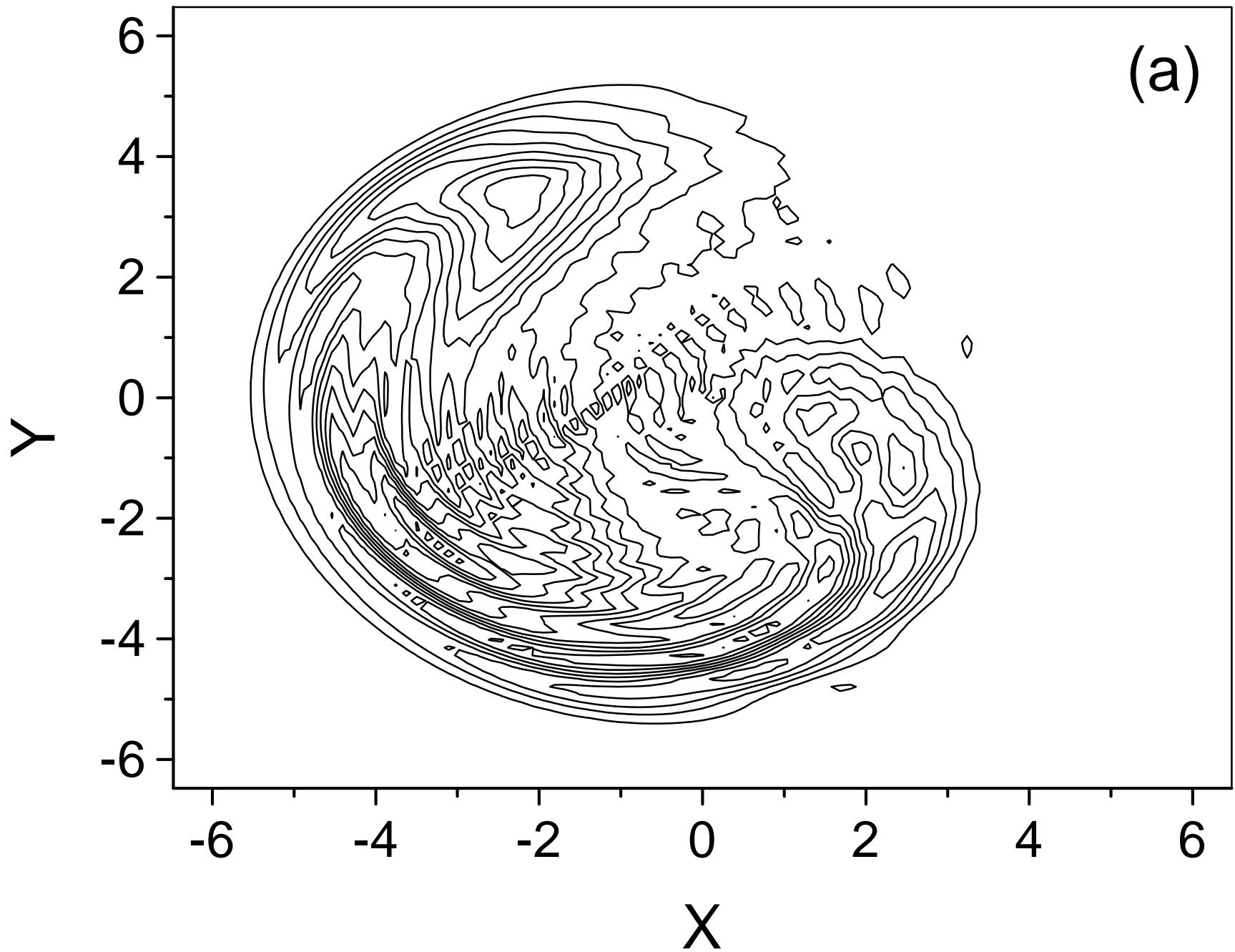
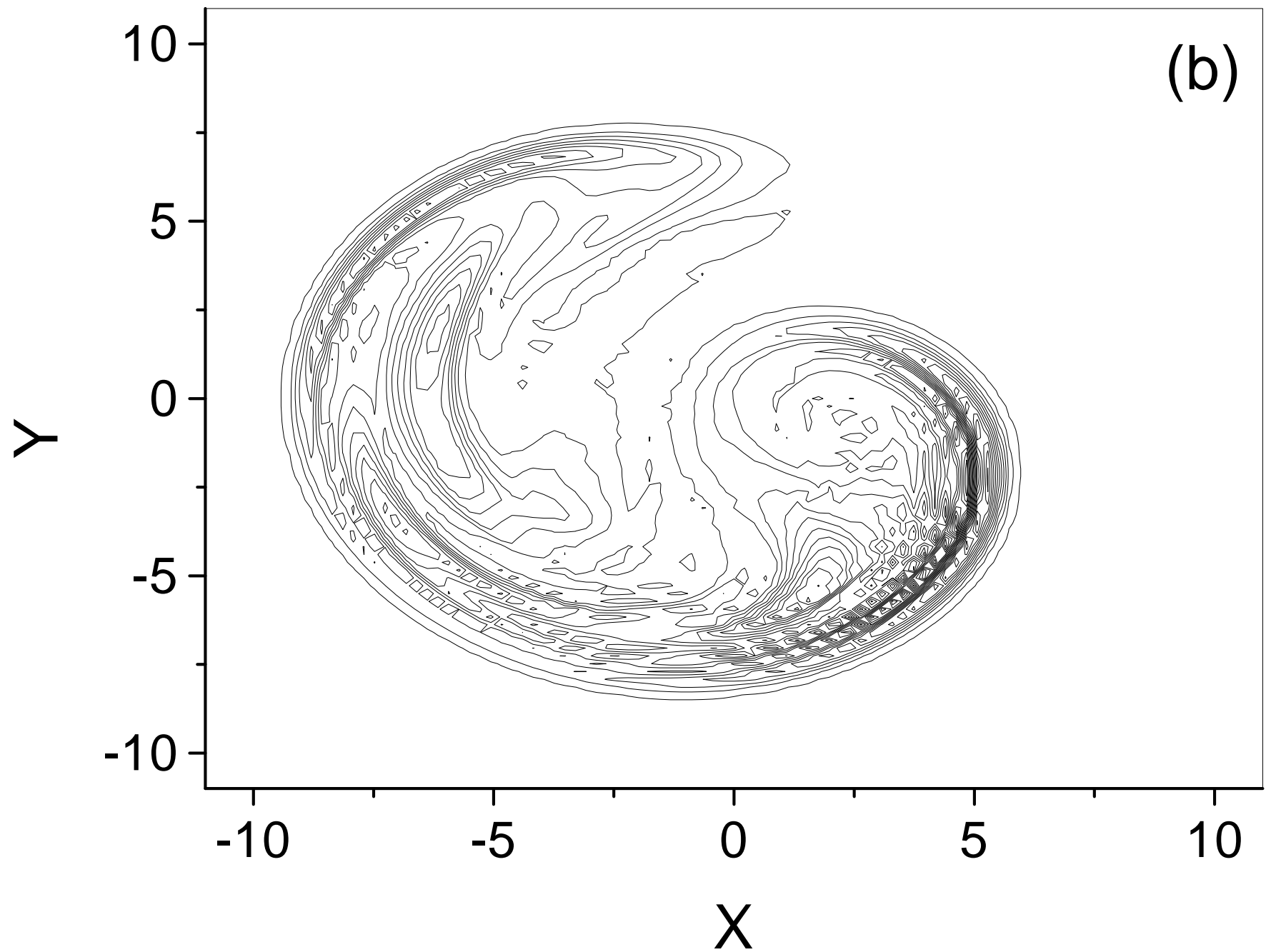
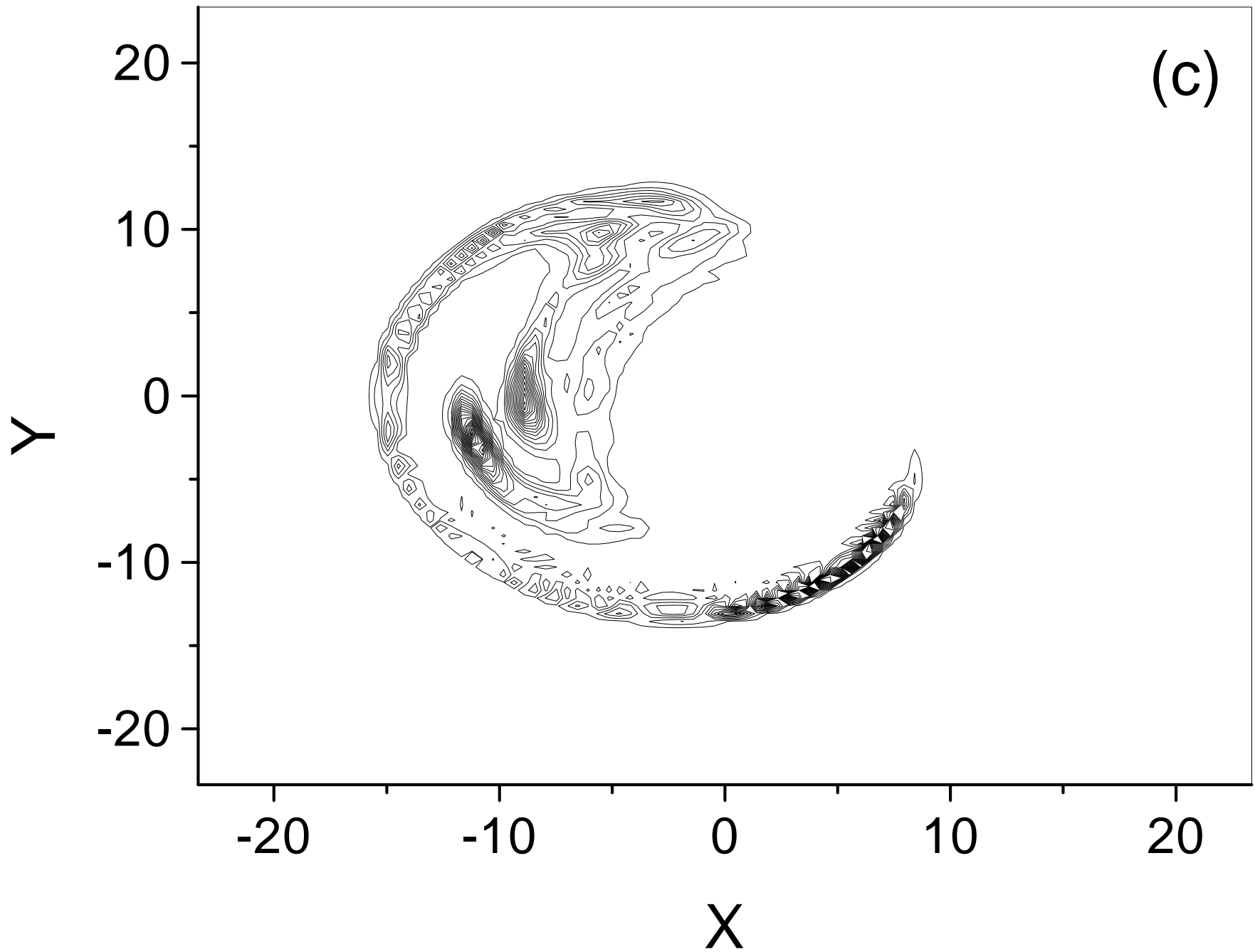
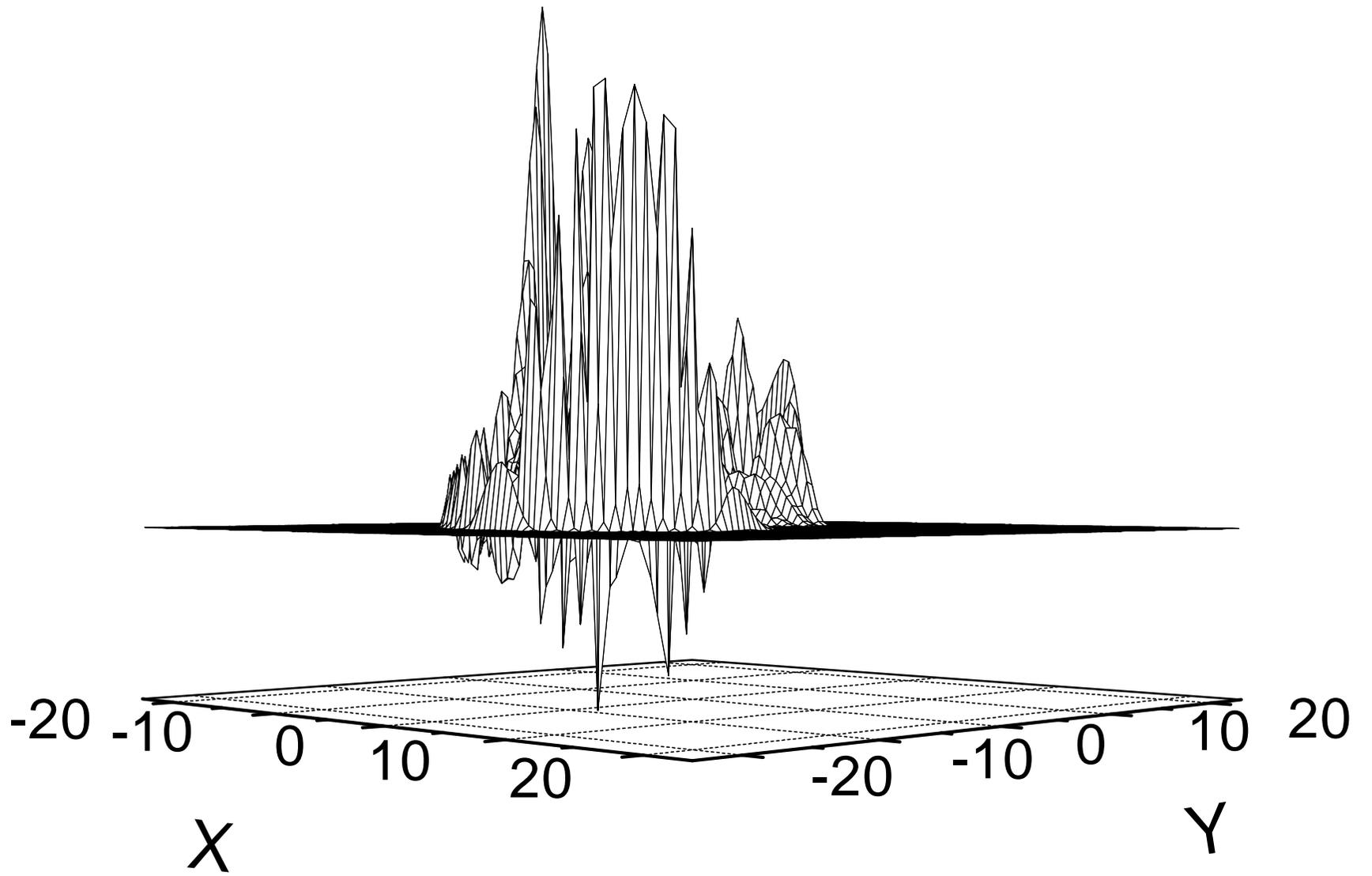
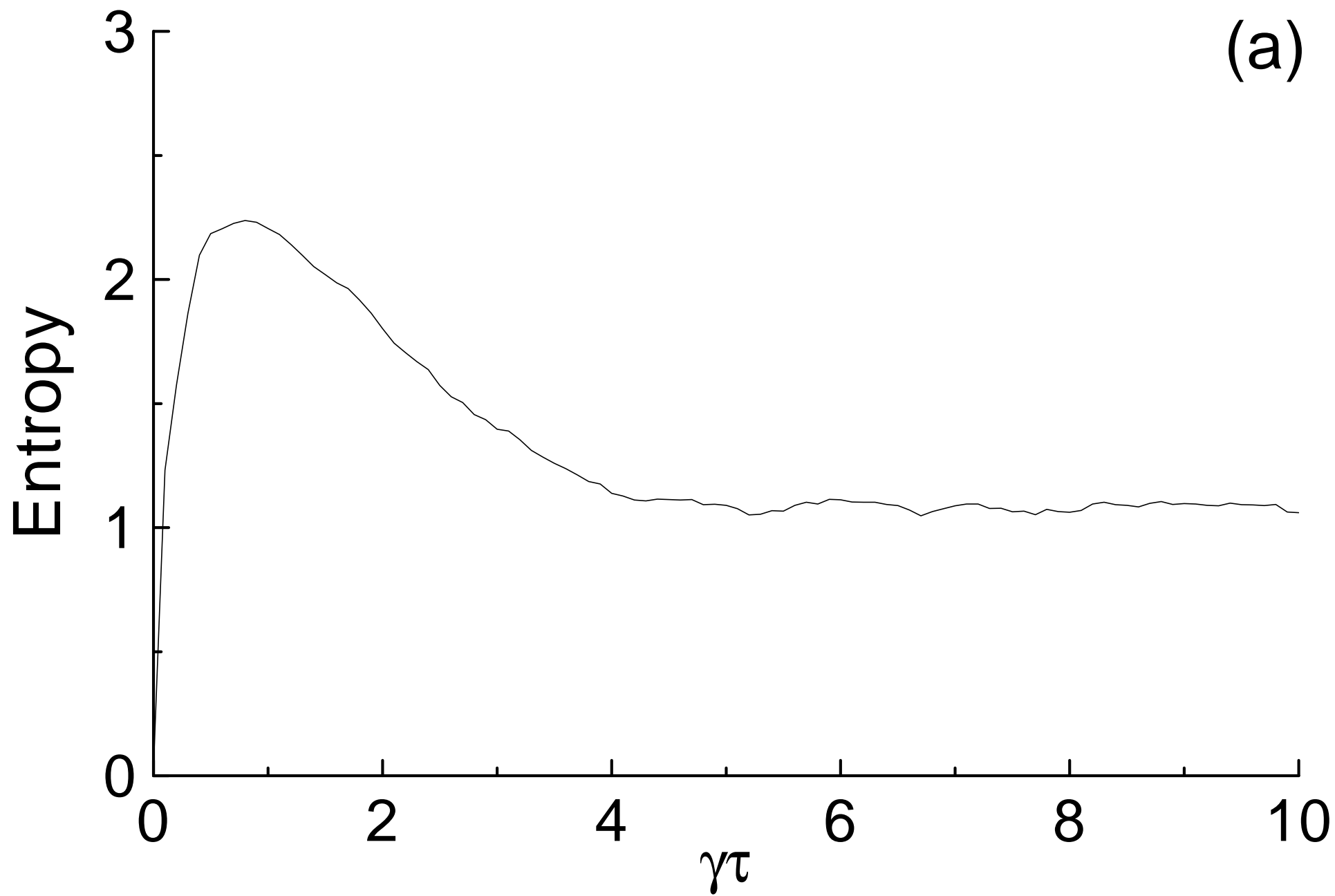
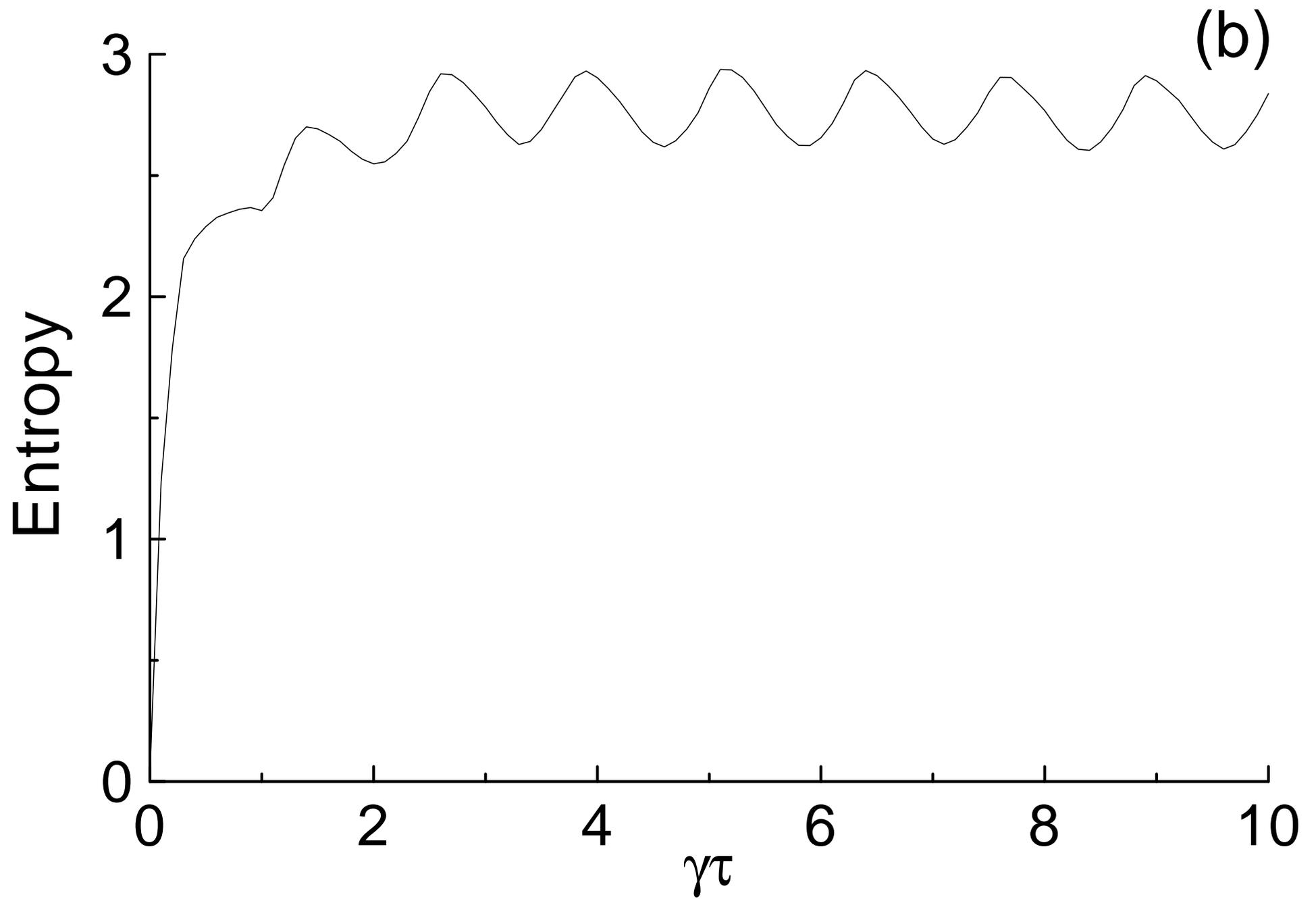
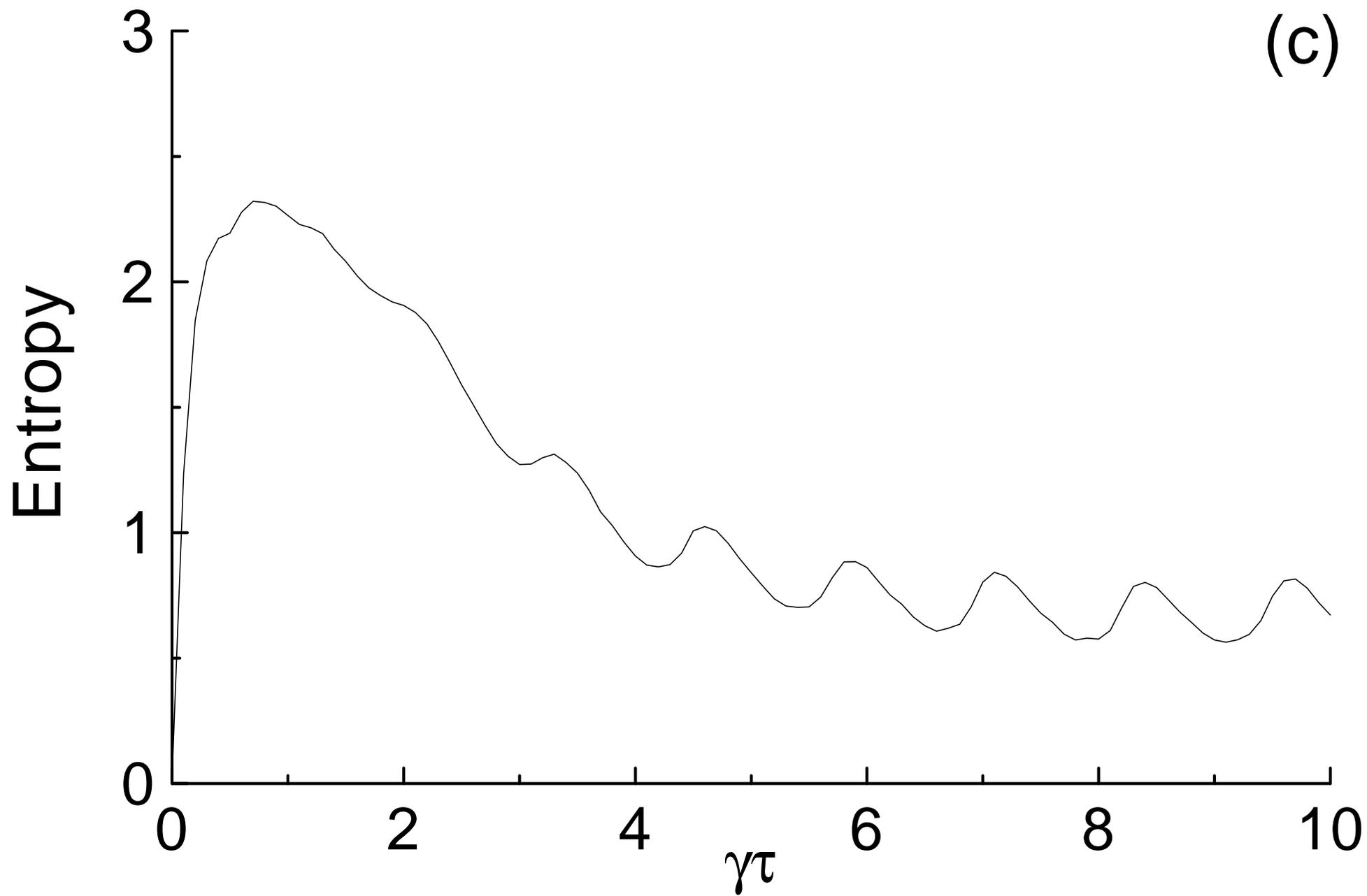
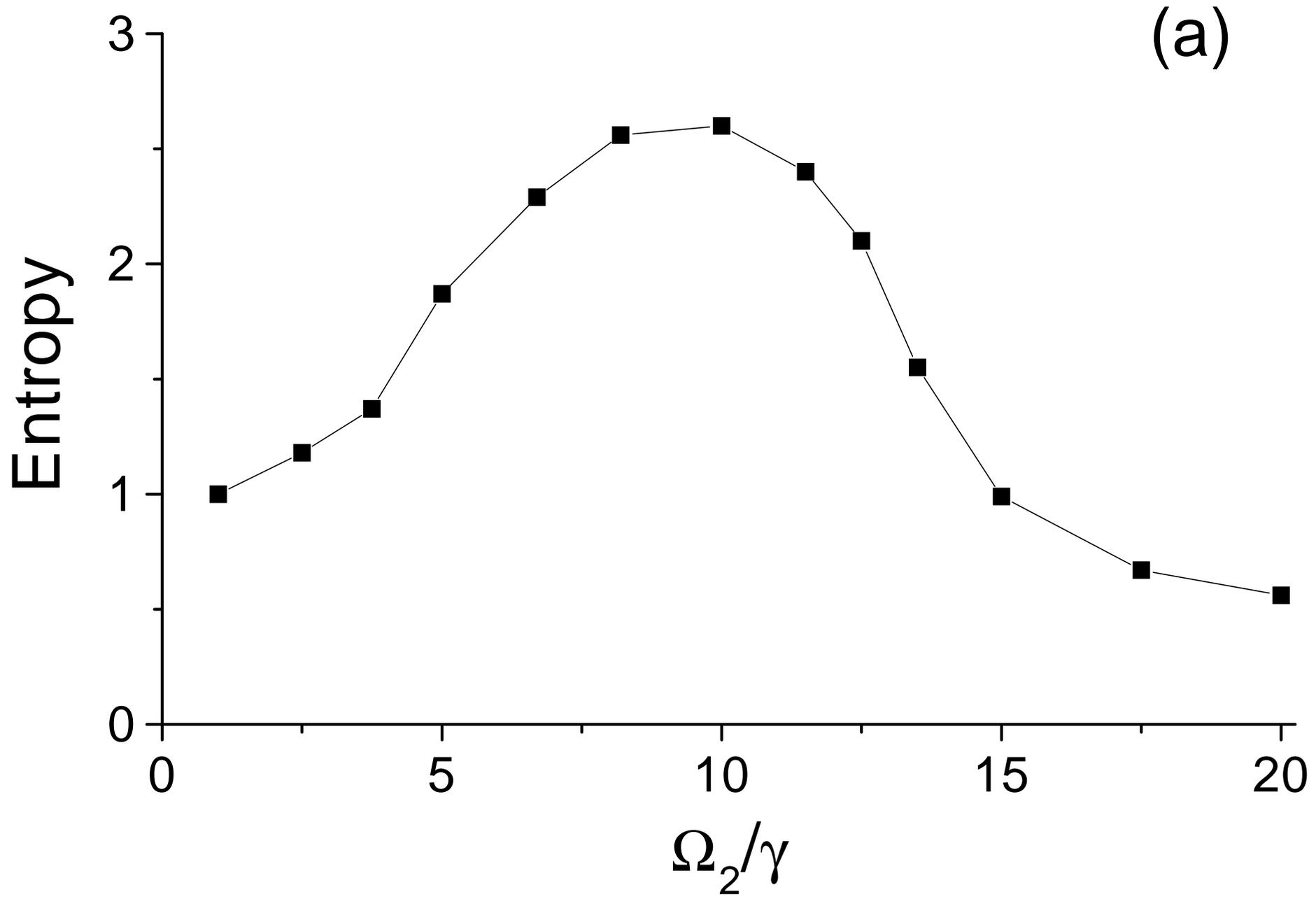
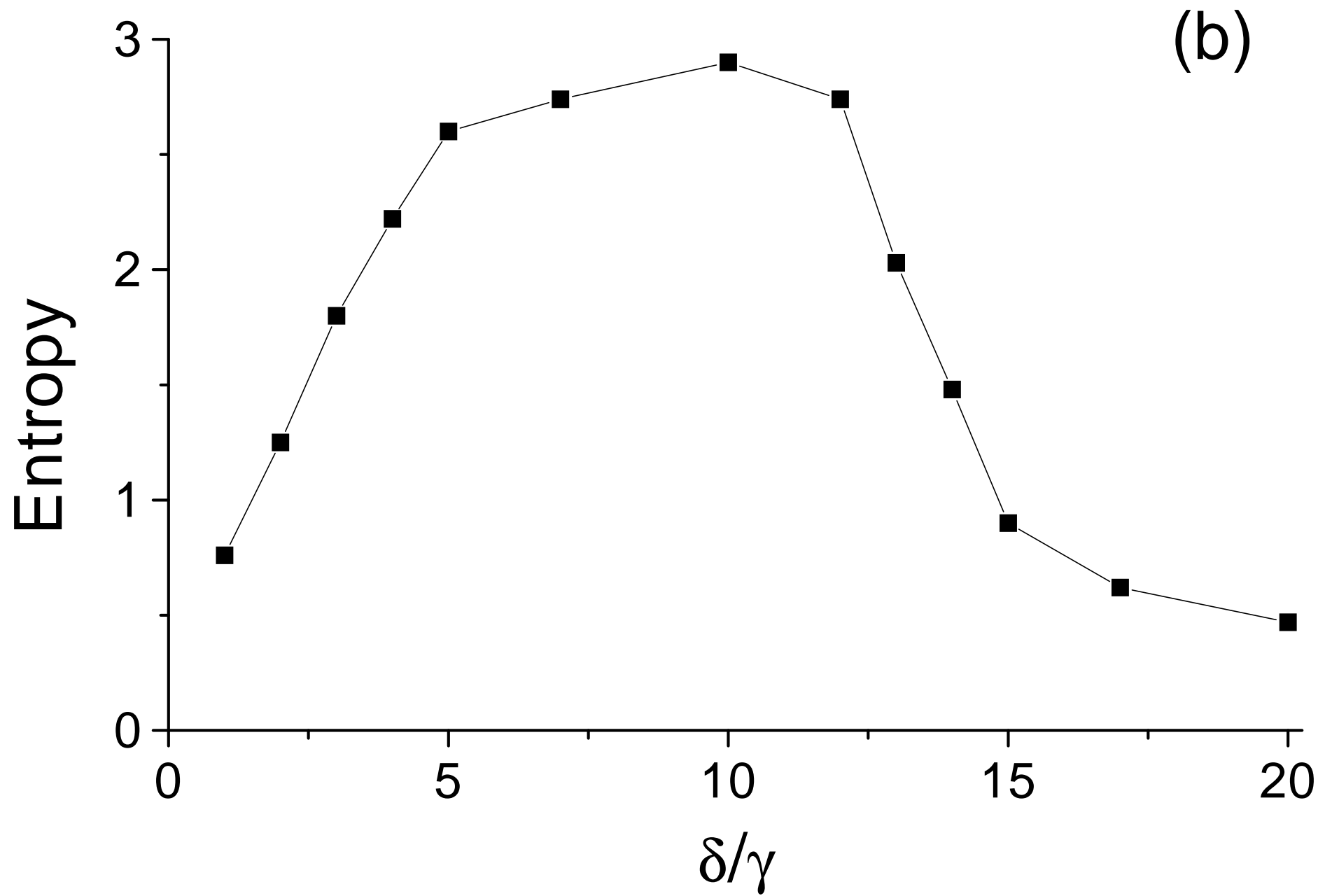
\end{document}